\documentclass[10pt, draftcls, onecolumn]{IEEEtran}

\usepackage[dvips,colorlinks,bookmarksopen,bookmarksnumbered,citecolor=red,urlcolor=red]{hyperref}

\newcommand\BibTeX{{\rmfamily B\kern-.05em \textsc{i\kern-.025em b}\kern-.08em
T\kern-.1667em\lower.7ex\hbox{E}\kern-.125emX}}

\usepackage{kbordermatrix}
\usepackage[ieee]{jphmacros2e}

\usepackage{pdfsync,srcltx}
\usepackage{amsmath}
\usepackage{amsfonts}
\usepackage{amssymb}
\usepackage{graphicx,subfigure,psfrag}
\usepackage{url,ifthen}

\newcommand{\R}{\mathbb{R}}

\newcommand{\N}{\mathbb{N}}
\newcommand{\Z}{\mathbf{Z}}
\renewcommand{\L}{\mathcal{L}}
\newcommand{\G}{\mathbf{G}}
\newcommand{\V}{\mathbf{V}}
\renewcommand{\E}{\mathbf{E}}
\newcommand{\scr}{\mathcal}

\newtheorem{thm}{Theorem}

\begin{document}

\title{\LARGE Decentralized control of large vehicular formations: stability margin and sensitivity to external disturbances}

\author{He Hao,  Prabir Barooah
\thanks{He Hao. He Hao and Prabir Barooah are with Department of Mechanical and Aerospace Engineering,
        University of Florida, Gainesville, FL 32611, USA. Email: {\small {hehao,pbarooah}@ufl.edu}. This work was supported by the National Science Foundation through Grant CNS-0931885 and by the Institute for Collaborative Biotechnologies through grant DAAD19-03-D-0004.
}
}

\maketitle

\begin{abstract}
	We study the stability and robustness of large-scale vehicular formations, in which each vehicle is modeled as a double-integrator. Two types of information graphs are considered: directed trees and undirected graphs. We prove stability of the formation with arbitrary number of vehicles for linear as well as a class of nonlinear controllers. In the case of linear control, we provide quantitative scaling laws of the stability margin and sensitivity to external disturbances ($H_\infty$ norm)  with respect to the number of vehicles $N$ in the formation. It is shown that the formation with directed tree graph achieves size-independent stability margin but suffers from high algebraic growth of initial errors. The stability margin in case of the undirected graph decays to $0$ as at least $O(1/N)$. In addition, we show that the sensitivity to external disturbances in directed tree graphs is \emph{geometric} in $k$ where $k\leq N$ is the number of generations of the directed tree, while that of the undirected graph is only \emph{quadratic} in $N$. In particular, for 1-D vehicular platoons, we obtain precise formulae for the $H_\infty$ norm of the transfer function from the disturbances to the position errors. It is shown that the $H_\infty$ norm scales as $O(\alpha^N) (\alpha>1)$ for predecessor-following architecture, but only as $O(N^3)$ for symmetric bidirectional architectures. For a class of nonlinear controllers, numerical simulations show that the transient response due to initial errors and sensitivity to external disturbances are improved considerably for the formation with directed tree graphs. However, by using the nonlinear controller considered, little improvement can be made for that with undirected graphs.
\end{abstract}


\section{Introduction}
 Distributed control of multi-agent systems has spurred an extensive interest in the control community because of its wide range of applications such as automated highway system~\cite{JH_MT_PV_CSM:94}, collective behavior of bird flocks and animal swarms~\cite{okubo_flock}, and coordination of aerial, ground, and autonomous vehicles for surveillance, energy savings, mine sweeping etc.~\cite{tanner2007decentralized,DJ_WB_MP_EW_AIAA:02,minesweeping}. A typical issue in distributed control is that as the number of agents in the system increases, the performance of the closed-loop degrades progressively. A classical problem in which this progressive loss of performance was observed is the coordination of a platoon of vehicles moving in a straight line, see~\cite{JH_MT_PV_CSM:94,stringstability:96,Seiler_disturb_TAC:04,zhang1999using,darbha1994comparison} and references therein. The goal in the platoon control problem is to maintain a constant separation between pairs of nearby vehicles using a control law that is distributed, meaning that each vehicle can only use measurements of relative position and/or velocity with respect to its (at most two) neighbors that can be obtained with on-board sensors. 

Two decentralized control architectures that are commonly examined in the literature are predecessor-following and bidirectional architectures. In the \emph{predecessor-following} architecture, the control action on each vehicle only depends on the relative information from its immediate predecessor, i.e. the vehicle in front of it.  In the bidirectional architecture, the control depends on the relative information from both its immediate predecessor and follower. Within the bidirectional architecture, the most commonly analyzed case is the \emph{symmetric bidirectional} architecture, in which the control at a vehicle depends on the information from both of its neighbors \emph{equally}. 

It has been established that the predecessor-following architecture suffers from high sensitivity to disturbances with linear control, see for instance~\cite{peppard_string,chu_platoon} and references therein. High sensitivity to external disturbance is typically referred to as string instability~\cite{stringstability:96,klinge_string:09} or slinky-type effect~\cite{zhang1999using,darbha1994comparison}. Seiler \emph{et. al.} showed that with linear control, string instability with the predecessor-following architecture is independent of the design of the controller, but a fundamental artifact of the architecture~\cite{Seiler_disturb_TAC:04}. String instability can be ameliorated by non-identical linear controllers but at the expense of the control gains increasing without bound as the number of the vehicles increases~\cite{darbha1994comparison,Khatir}. It was later shown in~\cite{Seiler_disturb_TAC:04,PB_JH_CDC:05,lestas_scalability} that the symmetric bidirectional architecture, also suffers from poor sensitivity to external disturbances. 

The degeneration of performance occurs not only in 1-D vehicular platoons, but also in vehicular formations moving in higher dimensional space. Mesh stability, which is a generalization of string stability is proposed in~\cite{AP_PS_KH_TAC:02} to study look-ahead interconnected systems. In~\cite{SKY_SD_KRR_TAC:06,SD_PP_IJES:10}, it is shown that 
if the information graph used is connected and undirected, the maximum error can be made independent of the size of the formation only if there is at least one vehicle communicating with at least $O(N)$ other vehicles in the formation. However, it was later shown in~\cite{SD_PP_IJES:10}, if there are at least two vehicles communicating to $O(N)$ other vehicles in the formation, the system will be unstable if its size is beyond a critical value. Moreover, if the information graph has bounded degrees, the maximum error due to disturbances will be at least $O(N)$.

Although a rich literature exists on sensitivity to disturbances for predecessor-following and symmetric bidirectional architectures with linear control, to the best of our knowledge, a precise quantitative measure  and comparison of sensitivity to disturbances of these two architectures is lacking. Moreover, most of the work on formation control of platoons has been limited to linear control laws, while little is known about nonlinear control. Nonlinear terms in the closed loop dynamics may arise from either purposefully designed nonlinear control laws (if beneficial) or unavoidable non-linearities in the vehicle dynamics, such as actuator saturations. Both of these cases can be analyzed by considering linear plant dynamics and nonlinear controllers.

In this paper we examine the stability and robustness (sensitivity to external disturbances) of vehicular formations with linear as well as a class of nonlinear controllers, for two types of information graphs: directed trees and undirected graphs. The 1-D vehicular platoons with predecessor-following and symmetric bidirectional architectures are special cases of the formations we considered. Each vehicle in the formation is modeled as  a fully actuated point mass (double-integrator).  This is a commonly used model for vehicle dynamics in studying vehicular formations, which results from feedback linearization of actual non-linear vehicle dynamics~\cite{darbha1994comparison,feedback_linearization}. A few authors have used first order kinematic models by ignoring vehicle inertia. However, it is pointed out in~\cite{SD_PP_IJES:10} that kinematic models fail to reproduce phenomena such as high sensitivity to disturbance that are exhibited by large platoons with kinetic models of vehicles. We prove stability of the closed loop with an arbitrary number of vehicles for a class of non-linear controllers, where the control gain functions satisfy certain sector conditions. The difference between the transient responses with different type of graphs in case of linear control is explained by the expressions we derive for the least stable eigenvalue of the closed-loop state matrix and its multiplicity. In particular, we show that  the formation with directed tree graph has a larger stability margin compared to that with undirected graph: constant vs. $O(1/N)$. However, it suffers from algebraic growth of initial conditions due to the high multiplicity of the least stable eigenvalue. For the non-linear control, we study the transient performance through  numerical simulations, which indicate that the transient response of the formation with directed tree graph can be improved significantly by using a saturation-type nonlinearity in the control. 

Next, we examine the closed-loop's performance in terms of the sensitivity to external disturbances. Specifically, we examine the \emph{amplification factor}, defined as the $L_2$ gain from the disturbances acting on all the vehicles to their position tracking errors. In case of linear controllers, we prove that the  amplification factor, which becomes a $H_\infty$ norm, grows as $O(\alpha^k)$ ($\alpha>1$) for directed tree graphs where $k$ is the number of generations of the directed tree,  but only as  $O(N^3)$ for the that with undirected graphs. For 1-D vehicular platoons, $k=N$ in the predecessor-following architecture. This establishes a precise comparison between the symmetric bidirectional and predecessor-following architectures in terms of robustness to disturbances. In addition, we show that the peak frequency for the predecessor-following architecture is $O(1)$ while for the symmetric bidirectional architecture it is $O(1/N)$. Establishing such scaling laws in case of non-linear systems has so far been proved intractable.  We therefore study the response in the non-linear case through extensive numerical simulations, with both sinusoidal and random disturbances as inputs, and estimate performance metrics from simulation data. We observe from these studies that,  within the predecessor-following architecture, a nonlinear controller with a saturation-type nonlinearity performs better than the corresponding linear one. In the symmetric bidirectional architecture, the difference between the linear and nonlinear controller's performance is not that significant, though these conclusions are valid only for the specific controllers we investigated.

In terms of the stability analysis with non-linear controllers, our work closely parallels that of~\cite{munz2011robust}. However,  the results of~\cite{munz2011robust} only  considers absolute velocity feedback while we consider relative velocity feedback.  Furthermore, the assumption of symmetry made in~\cite{munz2011robust} precludes the predecessor-following architecture from their formulation.  In terms of sensitivity to external disturbances with linear control, our work is related to~\cite{SD_PP_IJES:10,SKY_SD_KRR_TAC:06},~\cite{veerman_stability} and~\cite{bamjovmitCDC08}. The discussion on~\cite{SD_PP_IJES:10,SKY_SD_KRR_TAC:06} have been stated above. In~\cite{veerman_stability}, it was shown that for a 1-D vehicular platoon, a disturbance sensitivity metric grows linearly in $N$ for the symmetric bidirectional case.  Veerman~\cite{veerman_stability} also examined an asymmetric bidirectional architecture, though not the predecessor-following one, and showed that amplification of errors grow exponentially in $N$ in that case. Bamieh \text{et al.}~\cite{bamjovmitCDC08} obtained scaling laws of certain $H_2$ norms from disturbance to outputs that quantify a number of performance measures such as local error, long-range deviation etc. In this paper we consider the $H_\infty$ norm metric.

The rest of this paper is organized as follows. Section~\ref{sec:problem} presents the graph theory and the problem statement. The stability analysis appear in Section~\ref{sec:stability}. The results on sensitivity to external disturbances are stated in Section~\ref{sec:h8norm}. The paper ends with a conclusion in Section~\ref{sec:conc}.

\section{Graph theory and Problem statement}\label{sec:problem}
\subsection{Graph Theory}\label{sec:graph}
In this paper, we use an \emph{information graph} to model the interaction topology between controlled vehicles. An \emph{information graph} is a  graph   $\G = (\V,\E)$ with vertex set $\V$ and edge set $\E$. The set of edges $\E \subset \V \times \V$ specify the information flow used to compute their local control actions. We consider two types of information graph: \emph{directed} graph and \emph{undirected} graph. In a \emph{directed} graph, we use $(i,j)$ to represent a directed edge from $i$ to $j$, in which $i$ is called the parent of $j$ and $j$ is called the child of  $i$. The set of neighbors of $i$ is defined as $\scr{N}_i \eqdef \{j \in \V: (j, i)\in \E\}$. In particular, we restrict ourselves to a special class of directed graphs, \emph{directed trees}. A \emph{directed tree} is a directed graph which would be a tree if the directions on the edges were ignored.  We say a directed tree has a spanning tree if every node, except one special node without any parent, which is called the root, has exactly one parent, and the root can be connected to any other nodes through paths. For a directed tree graph who has a spanning tree,  we call the nodes whose parent is the root (reference vehicle) as the first generation, the nodes whose parent is in the first generation as the second generation  and so forth. Without loss of generality, we assume there are $n_k$ nodes in the $k$-th generation, and the nodes in it are labeled as $\{n_{k-1}+1,\cdots, n_{k-1}+n_k\}$ by identifying $n_0=0$. 

In an \emph{undirected} graph, the edges have no direction, i.e. $i \in \scr{N}_j$ if and only if $j \in \scr{N}_i$. We say an undirected graph has a spanning tree if there is a tree composed of all the vertices and some (or perhaps all) of the edges. For an undirected graph, it has a spanning tree if and only if it's connected. In both directed and undirected graphs, the \emph{graph Laplacian} $L$ is defined by $L_{ij}=-1$ if $e_{ji} \in \E$ and $L_{ii}=-\sum_{k} L_{ik}$. The \emph{grounded (Dirichlet) graph Laplacian} $L_g$ of the information graph is obtained by removing from $L$ the row and column corresponding to the grounded node (reference vehicle). 

\subsection{Problem Statement}
We consider the formation control of $N+1$ homogeneous vehicles which are moving in  Euclidean space. For ease of exposition, we only consider one dimension of the translation motion. The analysis is also applicable to all three dimensions, as long as the dynamics of each of the coordinates of a vehicle's
position are decoupled and each coordinate can be independently controlled. Under this \emph{fully actuated} assumption, the closed loop dynamics for each coordinate of the position can be independently studied; see~\cite{HH_PB_PM_TAC:11,SD_PP_IJES:10} for examples. The position of the $i$-th vehicle is denoted by  $p_i$ and each vehicle is modeled as a double integrator:
\begin{align}\label{eq:vehicle-dynamics}
	\ddot{p}_i= u_i+w_i, \quad i\in \{1,2,\cdots,N\},
 \end{align}
where $u_i$ is the control input, and $w_i$ is the external disturbance. This is a commonly used model for vehicle dynamics in studying vehicular formations, which results from feedback linearization of actual non-linear vehicle dynamics~\cite{darbha1994comparison,feedback_linearization}.
\begin{figure}
	  \psfrag{O}{$O$}
	  \psfrag{o}{$0$}
	  \psfrag{l}{$1$}
	  \psfrag{X}{$X$}
	  \psfrag{x}{$x$}
	  \psfrag{d1}{\scriptsize$\Delta_{(0,1)}$}
	  \psfrag{d2}{\scriptsize$\Delta_{(N-1,N)}$}
	   \psfrag{v}{$v^{*}\;t$}
	   \psfrag{d}{\scriptsize $1/N$}
	   \psfrag{0}{\scriptsize $0$}
	   \psfrag{1}{\scriptsize $1$}
	   \psfrag{2}{\scriptsize $2$}
	   \psfrag{N-1}{\scriptsize $\ N-1$}
	   \psfrag{N}{\scriptsize $N$}
	   \psfrag{Di}{\scriptsize Dirichlet}
	   \psfrag{Ne}{\scriptsize \ Neumann}
\centering
\subfigure[Vehicular platoon]{\includegraphics[scale = 0.35]{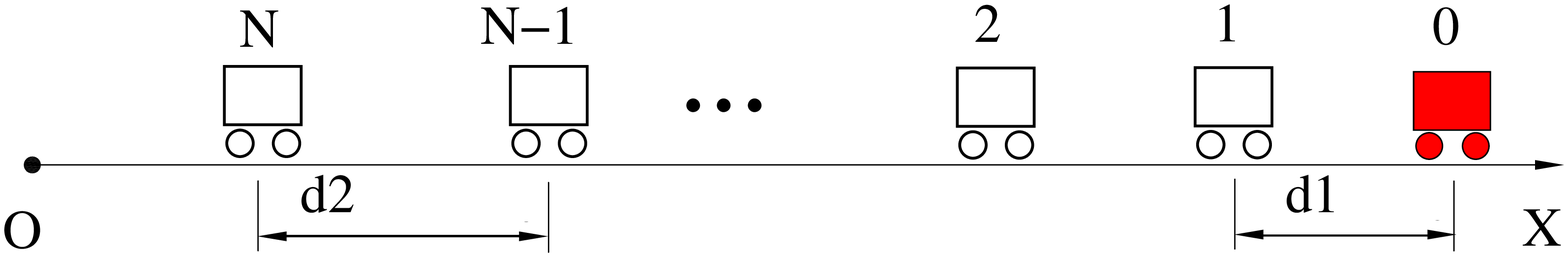}} \\
\subfigure[Directed information graph]{\includegraphics[scale = 0.2]{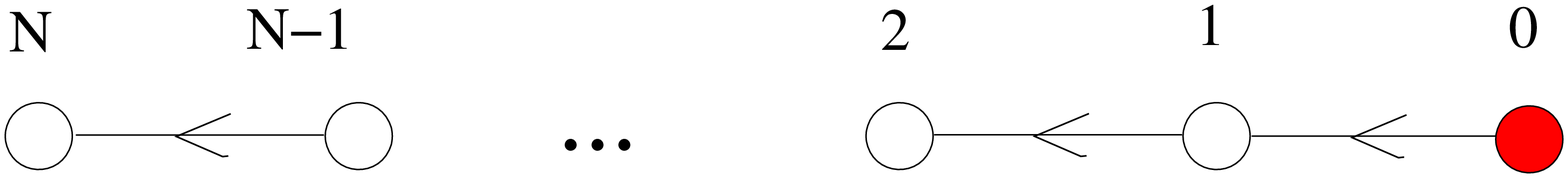}} \quad \quad \quad \quad
\subfigure[Undirected information graph]{\includegraphics[scale = 0.2]{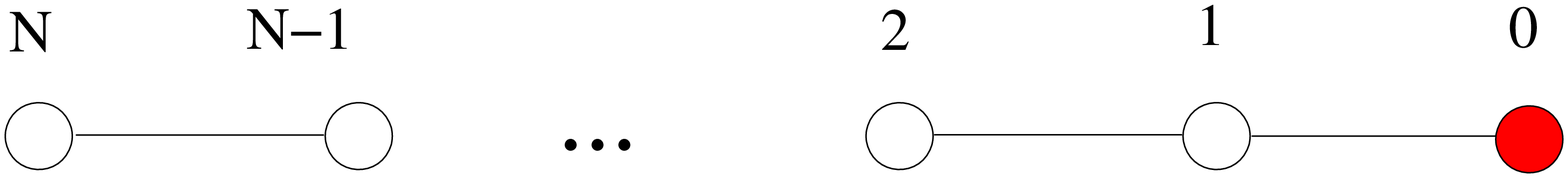}} 
\caption{Desired geometry and its information graphs of a vehicular platoon with $N$ vehicles and $1$ "reference vehicle", which are moving in 1D Euclidean space. The filled vehicle in the front of the platoon represents the reference vehicle, it is denoted by "$0$".}
\label{fig:fig1}
\end{figure}

The control objective is that vehicle maintain a rigid formation geometry while following a desired trajectory. The desired geometry of the formation is specified by the \emph{desired gaps} $\Delta_{(i,j)}$ for $i,j \in\{1,\cdots,N\}$, where $\Delta_{(i-1,i)}$ is the desired value of $p_{i}(t) - p_{j}(t)$. The desired inter-vehicular gaps $\Delta_{(i,j)}$'s are positive constants and they have to be specified in a mutually consistent fashion, i.e. $\Delta_{(i,k)}=\Delta_{(i,j)} +\Delta_{(j,k)}$ for every triple $(i, j, k)$. The desired trajectory of the formation is provided in terms of a reference vehicle. Since we are interested in translational maneuvers of the formation, we assume there is only one reference vehicle, which is denoted by index $0$. The trajectory of the reference vehicle is denoted by $p^*_0(t)$ and its dynamics is independent of the other vehicles.  The desired trajectory of the $i$-th vehicle, $p^*_i(t)$, is given by 
\begin{align}\label{eq:ref-p-star}
p^*_i(t) =p^*_0(t)-\Delta_{(0,i)}.  
\end{align}

In this paper, we consider the following \emph{decentralized} control laws, where the control action on each vehicle only depends on the relative position and velocity information from its neighbors in the information graph:
\begin{align}\label{eq:u}
	u_{i} = - \sum_{j \in \scr{N}_i} \Big (f (p_{i}-p_{j}+\Delta_{(j,i)}) +g(\dot{p}_{i}-\dot{p}_{j}) \Big).
\end{align}
In the above controller, we assume the possibly nonlinear functions $f, g:\R \to \R$ are odd functions, which are smooth enough to guarantee the existence of solution of the coupled ODEs. Note that the information needed to compute the control action can be easily accessed by on-board sensors such as radar and doppler sensors.

Moreover, in this paper, we are particularly interested in a special class of formations, i.e. vehicular platoons. Figure~\ref{fig:fig1} shows a pictorial representation of a vehicular platoon and its directed and undirected information graphs we considered in this paper. Corresponding to each type of the information graph (directed tree and undirected graph), we consider the following two \emph{decentralized} control architectures: 
\begin{enumerate}
\item \emph{Predecessor-following architecture}. The control action at the $i$-th vehicle depends on the relative measurements from its immediate front neighbor, which results in the following control law:
\begin{align}\label{eq:u-pred}
	u_{i} = &-f (p_{i}-p_{i-1}+\Delta_{(i-1,i)}) -g(\dot{p}_{i}-\dot{p}_{i-1}).
\end{align}
\item \emph{Symmetric bidirectional architecture}. The control action at the $i$-th vehicle depends equally on the relative  measurements from its immediate front and back neighbors, which results in the following control law:
\begin{align}\label{eq:u-bidi}
		u_{i} = &-f (p_{i}-p_{i-1}+\Delta_{(i-1,i)}) -g(\dot{p}_{i}-\dot{p}_{i-1})\notag \\&-f (p_{i}-p_{i+1}-\Delta_{(i,i+1)})-g(\dot{p}_{i}-\dot{p}_{i+1}), \notag \\
	u_{N} = &-f (p_{N}-p_{N-1}+\Delta_{(N-1,N)}) -g(\dot{p}_{N}-\dot{p}_{N-1}), 
\end{align}
where $i\in \{1,2,\cdots,N-1\}$. 
\end{enumerate}

For a vehicular formation which is desired to translate at a constant speed, we assume the reference trajectory is to be a constant velocity type, i.e., $p^*_0(t) = v_0t + c_0$ for some constants $v_0,c_0$. To facilitate analysis, we define the following  position tracking error:
\begin{align}\label{eq:state_error}
	\tilde{p}_i \  \eqdef \ p_i-p_i^*,
\end{align}
where $p_i^*$ is given in~\eqref{eq:ref-p-star}. The closed-loop dynamics of the vehicular formation can now be expressed as the following coupled-ODE model
\begin{align}\label{eq:coupled_ODE}
\ddot{\tilde{p}}_{i} = - \sum_{j \in \scr{N}_i} \Big (f(\tilde{p}_{i}-\tilde{p}_{j})+g(\dot{\tilde{p}}_{i}-\dot{\tilde{p}}_{j})\Big)+w_i,
\end{align}
where $i\in \{1,2, \cdots, N\}$. Note that $\tilde{p}_{0} (t) =\dot{\tilde{p}}_{0}(t) \equiv 0$ since the reference vehicle perfectly tracks its desired trajectory. The system can be expressed in the state space form:
\begin{align}\label{eq:ss-nl}
  \dot{ x} = \mathbf{f}(x, w),
\end{align}
where the state and disturbance vectors are defined as $x\eqdef [\tilde{p}_1, \dot{\tilde{p}}_1,\cdots, \tilde{p}_N, \dot{\tilde{p}}_N]^T$ and $w\eqdef[w_1,\cdots, w_N]^T$. The special case $f(z)=k_0z$ and $g(z)=b_0z$ (where $k_0>0, b_0>0$) in the above coupled-ODEs correspond to the case of \emph{linear control}.  In that case, the closed-loop can be represented as:
\begin{align}\label{eq:ss}
	\dot{x}=A x+Bw,
\end{align} 
where $A$ is the state matrix that depends on $k_0, b_0$ and $B$ is the input matrix with appropriate dimension.
\section{Stability Analysis}\label{sec:stability}

In this section, we present the stability analysis of the undisturbed vehicular formations  with both linear and  nonlinear controllers. For the linear case, we also derive formulae showing  the relationship between the \emph{stability margin} and the smallest and largest eigenvalues of its grounded graph Laplacian. The \emph{stability margin} is defined as the absolute value of the real part of the least stable eigenvalue of the state matrix $A$ in~\eqref{eq:ss}.  The  stability margin quantifies the system's convergence rate with respect to initial errors. For the case of non-linear controller, we provide sufficient conditions for asymptotic stability. Since convergence rates for non-linear systems are difficult to obtain analytically, we perform numerical simulations to study the convergence rate and transient performance with non-linear controllers and compare with that with linear controllers. In particular, we take 1-D vehicular platoon as examples. All simulations for studying transient performance correspond to the following scenario: we perturb the initial position  of the first vehicle in the platoon from its desired value and observe the  position tracking error of the last vehicle $\tilde{p}_N(t)$. In addition, for the convenience of comparison, we define the following energy measure of transient performance: 
\begin{align*}
	E \eqdef  \lim_{T\to\infty} \frac{1}{x_0^2} \int_0^T \tilde{p}_N^2(t)+ \dot{\tilde{p}}_N^2(t) \ d t,
\end{align*}
where  $x_0$ is the initial error of the first vehicle, i.e., $\tilde{p}_1(0)=x_0$.  We assume the above limit exits , i.e. the last vehicle has finite $\L_2$ energy. In numerical simulations, we use the following estimate of $E$,
\begin{align}\label{eq:energy}
	\hat{E} \eqdef  \frac{1}{x_0^2} \int_0^T \tilde{p}_N^2(t)+ \dot{\tilde{p}}_N^2(t) \ d t,
\end{align}
where $T$ is sufficiently large such that all the errors die out. We study through numerical simulations how $E$ scales with the number of vehicle $N$ and the initial error $x_0$.

\subsection{Stability analysis  with linear control}
With linear controller $f(z)=k_0z$, $g(z)=b_0z$ ($k_0>0,b_0>0$) in~\eqref{eq:u}, it's straightforward to see that the state matrix $A$ can be expressed in the following form,
\begin{align}\label{eq:matrix_relation}
 	A = I_N \otimes A_1 + L_g \otimes A_2,
\end{align}
where $\otimes$ denotes the Kronecker product, $I_N$ is the $N \times N$ identity matrix and $A_1, A_2$  are given as below
\begin{align}\label{eq:auxi_matrices}
	A_1= \begin{bmatrix}
		0 & 1 \\
		0 & 0 \\
	       \end{bmatrix}, \quad
	A_2= \begin{bmatrix}
		0 & 0 \\
		-k_0 & -b_0 \\
	       \end{bmatrix}.
\end{align}
The matrix $L_g$ is the grounded (Dirichlet) graph Laplacian of the vehicular formation with corresponding information graph. For example, the ground graph Laplacian for the 1-D vehicular platoon with the directed tree (predecessor-following) and undirected (symmetric bidirectional) information graphs are given as below:
\begin{align}\label{eq:redu_laplacian}
L^{(d)}_g=
	       \kbordermatrix{ & 1 & 2 & \cdots  & N-1 & N\\
                1 & 1 &  &  & &  \\
		2& -1 & 1 &  & &\\
		 \vdots & & \ddots&\ddots  &  &  \\
		N-1& & &  -1 &1 &\\
		N & & & & -1 & 1},\ 
L^{(u)}_g=
	\kbordermatrix{ & 1 & 2 & \cdots  & N-1 & N\\
                1 & 2 & -1 &  & &  \\
		2& -1 & 2 & -1 & &\\
		 \vdots & &\ddots &\ddots  & \ddots &  \\
		N-1& & &  -1 &2 &-1\\
		N & & & & -1 & 1}.
\end{align}

We now present a formula for the stability margin of the formation in terms of the eigenvalues of its grounded graph Laplacian. 
\begin{theorem}\label{thm:eig_relation}
	Consider a vehicular formation with linear controller,  whose state matrix $A$ is given in~\eqref{eq:matrix_relation}. If the information graph has a spanning tree, then the system is globally exponential stable and its  stability margin is given by 
\begin{align}\label{eq:conv_rate}
	S =
	\begin{cases}
		\frac{b_0 \lambda_1}{2},& \text{if } \lambda_N \leq 4k_0/b_0^2, \\ 
		\frac{2k_0}{b_0+\sqrt{b_0^2-4k_0/\lambda_N}}, & \text{if } \lambda_1 \geq 4k_0/b_0^2, \\
		\min \Big \{ \frac{b_0 \lambda_1}{2}, \frac{k_0}{b_0+\sqrt{b_0^2-4k_0/\lambda_N}} \Big\},& \text{otherwise.}
	\end{cases}
\end{align}
where $0<\lambda_1 \leq \cdots \leq \lambda_N$ are the eigenvalues of the grounded graph Laplacian $L_g$.
Furthermore, 
\begin{enumerate}
	\item With directed tree graph, the stability margin is $S^{(d)}=O(1)$ and the least stable eigenvalue of the closed-loop state matrix $A$ is $s_{\min}=\frac{-b_0+ \sqrt{b_0^2-4k_0}}{2}$, and this eigenvalue occurs with multiplicity $N$. 
	\item With undirected graph with bounded degree, when $N$ is large, the stability margin satisfies $S^{(u)} \leq \frac{c^*}{N}$ for some positive constant $c^*$.\frqed 
\end{enumerate}
\end{theorem}

\begin{proof-theorem}{\ref{thm:eig_relation}}
Our proof follows a similar line of attack as~\cite[Theorem 4.2]{veerman_stability}. From Schur's triangularization theorem, every square matrix is unitarily similar to an upper-triangular matrix. Therefore, there exists an unitary matrix $U$ such that
\begin{align*}
 	U^{-1} L_g U= L_u,
\end{align*}
where $L_u$ is an upper-triangular matrix, whose diagonal entries are the eigenvalues of $L$. We now do a similarity transformation on matrix $A$.
 \begin{align*}
 	\bar{A} \eqdef &(U^{-1} \otimes I_2) A (U \otimes I_2)=(U^{-1} \otimes I_2) (I_N \otimes A_1 + L_g \otimes A_2) (U \otimes I_2)= I_N \otimes A_1+ L_u \otimes A_2.
\end{align*}
It is a block upper-triangular matrix, and the block on each diagonal is $A_1+\lambda_{\ell} A_2$, where $\lambda_{\ell} \in \sigma(L_g)$, where $\sigma(\cdot)$ denotes the spectrum. Since similarity preserves eigenvalues, and the eigenvalues of a block upper-triangular matrix are the union of eigenvalues of each block on the diagonal, we have
\begin{align}\label{eq:eigen_relation}
 	\sigma(A) = \sigma(\bar{A})  = \bigcup_{\lambda_{\ell} \in \sigma(L_g)} \{ \sigma (A_1+\lambda_{\ell} A_2)\}
 = \bigcup_{\lambda_{\ell} \in \sigma(L_g)} \Big \{ \sigma  \begin{bmatrix}
		0 & 1 \\ -k_0\lambda_\ell & -b_0 \lambda_\ell	       \end{bmatrix}
	       \Big \}.
\end{align}
It follows now that the eigenvalues of $A$ are the roots $s$ of 
\begin{align}\label{eq:chara_eqn}
	s^2+\lambda_{\ell} b_0 s + \lambda_{\ell}  k_0=0.
\end{align}

We now prove that the eigenvalues $\lambda_\ell$'s are real and positive. For a directed tree graph who has a spanning tree, by the labeling method given in Section~\ref{sec:graph}, it's straightforward to see that the ground graph Laplacian $L_g^{(d)}$ is a lower triangular matrix whose diagonal entries are  all $1$'s. Thus all its eigenvalues are real and positive. For a connected, undirected graph, it follows from~\cite[Lemma 1]{barooah2006graph} that the ground graph Laplacian $L_g^{(u)}$ is positive definite. In addition, it's symmetric, so its eigenvalues are real and positive.  Since $k_0>0, b_0>0, \lambda_\ell>0$, we obtain from~\eqref{eq:chara_eqn} that all the eigenvalues of $A$ has negative real part. Therefore, the system is globally exponential stable. 

For each $\ell \in \{1,2,\cdots, N\}$, the two roots of the characteristic equations~\eqref{eq:chara_eqn} are denoted by $s^{\pm}_\ell$,
 \begin{align}\label{eq:eigenvalue_pos_vel}
	 s_\ell^{\pm} =-\frac{\lambda_{\ell} b_0}{2}\pm \frac{\sqrt{(\lambda_{\ell}b_0)^2-4\lambda_{\ell}k_0}}{2}.
 \end{align}
 The one that is closer to the imaginary axis is denoted by $s_\ell^+$, and is called the \emph{less stable} eigenvalue between the two. The \emph{least stable} eigenvalue is the one closet to the imaginary axis among them, it is denoted by $s_{\min}$.

Depending the  discriminant in~\eqref{eq:eigenvalue_pos_vel}, there are three cases to analyze:
\begin{enumerate}
\item If $\lambda_N \leq 4k_0/b_0^2$, then the discriminant in~\eqref{eq:eigenvalue_pos_vel} for each $\ell$ is non-positive, then recall that the stability margin $S$ is defined as the absolute value of the real part of the least stable eigenvalue, which yields 
\begin{align*}
	S=|Re(s_{\min})|=\frac{\lambda_1 b_0}{2}.
\end{align*}
\item If $ \lambda_1 \geq 4k_0/b_0^2$, then the discriminant in~\eqref{eq:eigenvalue_pos_vel} for each $\ell$ is non-negative, the \emph{less stable} eigenvalue can be written as
\begin{align*}
	s_\ell^+=-\frac{\lambda_{\ell} b_0-\sqrt{(\lambda_{\ell}b_0)^2-4\lambda_{\ell}k_0}}{2}= -\frac{2k_0}{b_0+\sqrt{b_0^2-4k_0/\lambda_\ell}}.
\end{align*}
The least stable eigenvalue is achieved by setting $\lambda_\ell=\lambda_N$, then have the convergence rate
\begin{align*}
	S=|Re(s_{\min})|=\frac{2k_0}{b_0+\sqrt{b_0^2-4k_0/\lambda_N}}.
\end{align*}
\item Otherwise, if the discriminant in~\eqref{eq:eigenvalue_pos_vel} is indeterministic, i.e. it's negative for small $\ell$ and positive for large $\ell$, then the stability margin is given by taking the minimum of the above two cases.  
\end{enumerate}
In the case of directed tree graph, we know that the eigenvalues of the grounded graph Laplacian $L_g^{(d)}$ with directed tree graph are $\lambda_1=\cdots=\lambda_N=1$. From~\eqref{eq:eigenvalue_pos_vel}, we see that the eigenvalues of the state matrix $A$ are $(-b_0 \pm \sqrt{b_0^2-4k_0})/2$ and they have multiplicity of $N$. They are independent of $N$, thus the stability margin $S^{(d)}$ is $O(1)$. 	For an connected, undirected information graph whose maximum degree is a function $q(N+1)$, where $N+1$ is the number of vehicles (including the reference vehicle) in the formation. It follows from Lemma 3.2 of~\cite{SKY_SD_KRR_TAC:06} that $\lambda_1 \leq q(N+1)/{N}$. If $q(N+1)$ is bounded,  it follows from the three cases discussed above that when $N$ is large, $S^{(u)} \leq c^*/N$ for some positive constant $c^*$.\frQED
\end{proof-theorem}

\begin{remark}
For the connected, undirected graph, the upper bound on $\lambda_1$ can be tightened if the information graph is known beforehand~\cite{SKY_SD_KRR_TAC:06,SKY_MSThesis}. In particular,  for 1-D vehicular platoons, we obtain from Theorem 2 of~\cite{yueh_tridiag} that the eigenvalues $\lambda_\ell$ of $L_g^{(u)}$ are given by
$\lambda_\ell=2-2\cos(\frac{(2\ell-1)\pi}{2N+1})=4\sin^2(\frac{(2\ell-1)\pi}{2(2N+1)})$. They are distinct. For large $N$ and small $\ell$, we use the following approximation $\lambda_\ell=\frac{(2\ell-1)^2\pi^2}{4N^2}$, it's not difficult to see that the discriminant in~\eqref{eq:eigenvalue_pos_vel} is negative. Therefore, the least stable eigenvalue is determined by setting $\ell=1$, it has multiplicity $1$. The smallest eigenvalue $\lambda_1$ is now $O(1/N^2)$,  we obtain the stability margin $S^{(u)}=O{(1/N^2)}$. \frqed
\end{remark}

Although stability guarantees that transients due to initial conditions decay to $0$ as $t \to \infty$, the convergence rate and transient performance depend quite strongly on the architecture. For a stable LTI system $\dot{x} = Ax$,  the \emph{stability margin} is an appropriate measure of this convergence rate.  It follows from Theorem~\ref{thm:eig_relation} that the stability margin of the formation with linear control decays to $0$ with increasing $N$ for undirected information graph, while it is independent of $N$ for directed tree graph. This makes the formation with directed tree graph architecture have faster convergence rate than that with undirected graphs, especially for large $N$. However, the large algebraic multiplicity of the least stable eigenvalue of the formation with directed trees will cause large algebraic growth of the initial errors before they decay to $0$. For example, in a directed tree graph with $k$ generations,  the transient is proportional to $t^k e^{Re(s_{\min})t}$. For a directed tree graph with bounded degree, $k \to \infty$ as $N\to \infty$. So for large $N$, the algebraic growth is obvious.  Corroboration through numerical simulations is provided in Section~\ref{sec:conv_rate}.

\subsection{Stability analysis with non-linear control}
The next results are on the stability of the formation with non-linear controllers. In the statements of the theorems that follow we say that a scalar function $f$ belongs to the sector $[\varepsilon, K]$ if $\varepsilon z^2 \leq zf(z) \leq K z^2, \forall \ z \in \R$, and it belongs to the sector $(0,\infty]$ if $ zf(z) >0, \ \forall \ z \neq 0$. 

\begin{theorem}\label{thm:stability_pred}
	Consider a vehicular formation with directed tree information graph. If it has a  spanning tree, and $f(z), g(z)$ satisfy the sector conditions $f \in [\varepsilon_1, K_1], g \in [\varepsilon_2, K_2]$, where $0<\varepsilon_1 \leq K_1 < \infty, 0<\varepsilon_2 \leq K_2 < \infty$, then the origin $x=0$ is globally asymptotically stable. \frqed 
\end{theorem}
\begin{theorem}\label{thm:stability_bidi}
	Consider a vehicular formation with undirected information graph. If it has a spanning tree, and $f(z), g(z)$ satisfy the sector conditions $f \in (0,\infty], g \in (0,\infty]$, then the origin $x=0$ is globally asymptotically stable. \frqed 
\end{theorem}

\begin{remark}
Note that the stabilities of the linear controllers are special cases of Theorem~\ref{thm:stability_pred} and Theorem~\ref{thm:stability_bidi}. However, the latter two theorems have no implications on the stability margin. Comparing the above two theorems, we notice that the requirement on the sector condition with directed information graph is stricter than that of undirected information graph. However, we should note that these sector conditions are only sufficient. \frqed
\end{remark}

The proof of Theorem~\ref{thm:stability_pred} will use the following proposition, whose proof is given in the appendix.
\begin{proposition}\label{prop:2nd-order-stab}
Consider the second order autonomous system $\dot{y}_1 = y_2, \dot{y}_2 = -f(y_1-u_1) - g(y_2-u_2)$, where $y_1,y_2,u_1,u_2 \in \R$ and the odd functions $f,g:\R \to \R$ lie in the sectors $f \in [\varepsilon_1, K_1], g \in [\varepsilon_2, K_2]$, where $0<\varepsilon_1 \leq K_1 < \infty, 0<\varepsilon_2 \leq K_2 < \infty$. The origin of the unforced system (with $u(t)=[u_1(t), u_2(t)]^T \equiv 0$) is globally exponentially stable (GES) and the system is input-to-state stable (ISS) with $u$ as the input. \frqed     
\end{proposition}

\begin{proof-theorem}{\ref{thm:stability_pred}}
	We observe that in a directed tree information graph who has a spanning tree, the dynamics of each node (vehicle) only depend on the position and velocity of its parent, but not other vehicles. For example, let $i$ be an arbitrary vehicle in the graph, we use $\tilde{p}_i^{(+)}$ to denote the position error of its parent, the closed-loop dynamics of  $i$ can be written as:
\begin{align*}
	\ddot{\tilde{p}}_i=-f(\tilde{p}_i-\tilde{p}_i^{(+)})-g(\dot{\tilde{p}}_i-\dot{\tilde{p}}_i^{(+)}).
\end{align*}
According to Proposition~\ref{prop:2nd-order-stab}, the origin of the above unforced dynamics (with $u(t)=[\tilde{p}_i^{(+)}, \dot{\tilde{p}}_i^{(+)}]^T \equiv 0$) is GES, and it's ISS with $u(t)$ as its input. 

We first consider the subsystem consisted of the root (reference vehicle) and the first generation, whose closed-loop dynamics can be written as below by using the fact $[\tilde{p}_0, \dot{\tilde{p}}_0]^T \equiv 0$,
\begin{align*}
		\begin{cases}
		\ddot{\tilde{p}}_1&=-f(\tilde{p}_1)-g(\dot{\tilde{p}}_1),\\
				  &\vdots\\
		\ddot{\tilde{p}}_{n_1}&=-f(\tilde{p}_{n_1})-g(\dot{\tilde{p}}_{n_1}).
	\end{cases} \quad \Rightarrow \quad
	x^{(n_1)}=\mathbf{f}_{n_1}(x^{(n_1)}),
\end{align*}
where $n_1$ is the number of nodes in the first generation and $x^{(n_1)}=[\tilde{p}_1,\dot{\tilde{p}}_1,\cdots,\tilde{p}_{n_1},\dot{\tilde{p}}_{n_1}]^T$. We notice from Proposition~\ref{prop:2nd-order-stab} that the origin of each dynamics in the above equation is GES, thus the origin of the cascade system $x^{(n_1)}=\mathbf{f}_{n_1}(x^{(n_1)})$ is also GES. We next consider the subsystem consisted of the root and the first two generations. Its closed-loop dynamics can be written as:
\begin{align*}
	\begin{cases}
			\ddot{\tilde{p}}_1&=-f(\tilde{p}_1)-g(\dot{\tilde{p}}_1),\\
				  &\vdots\\
		\ddot{\tilde{p}}_{n_1}&=-f(\tilde{p}_{n_1})-g(\dot{\tilde{p}}_{n_1}),\\
		\ddot{\tilde{p}}_{n_1+1}&=-f(\tilde{p}_{n_1+1}-\tilde{p}_{n_1+1}^{(+)})-g(\dot{\tilde{p}}_{n_1+1}-\dot{\tilde{p}}_{n_1+1}^{(+)}),\\
				  &\vdots\\
				  \ddot{\tilde{p}}_{n_1+n_2}&=-f(\tilde{p}_{n_1+n_2}-\tilde{p}_{n_1+n_2}^{(+)})-g(\dot{\tilde{p}}_{n_1+n_2}-\dot{\tilde{p}}_{n_1+n_2}^{(+)}).
	\end{cases} \Rightarrow
	x^{(n_1+n_2)}=\mathbf{f}_{n_1+n_2}(x^{(n_1+n_2)}),
\end{align*}
where $x^{(n_1+n_2)}=[\tilde{p}_1,\dot{\tilde{p}}_1,\cdots,\tilde{p}_{n_1+n_2},\dot{\tilde{p}}_{n_1+n_2}]^T$. Notice that the parents of the nodes in the second generation are a subset of the nodes in the first generation, i.e. $\{\tilde{p}_{n_1+1}^{(+)}, \dot{\tilde{p}}_{n_1+1}^{(+)},\cdots,\tilde{p}_{n_1+n_2}^{(+)},\dot{\tilde{p}}_{n_1+n_2}^{(+)}\}$ is a subset of $\{\tilde{p}_1,\dot{\tilde{p}}_1,\cdots, \tilde{p}_{n_1},\dot{\tilde{p}}_{n_1}\}$. The above dynamics can be divided into two parts:
\begin{align}\label{eq:cas}
	x^{(n_1+n_2)}=\mathbf{f}_{n_1+n_2}(x^{(n_1+n_2)}), \quad \Rightarrow \quad 
	\begin{cases}
		x^{(n_1)}&=\mathbf{f}_{n_1}(x^{(n_1)}),\\
		x^{(n_2)}&=\mathbf{f}_{n_2}(x^{(n_2)}, x^{(n_1)}),
	\end{cases}
\end{align}
where $x^{(n_2)}=[\tilde{p}_{n_1+1},\dot{\tilde{p}}_{n_1+1},\cdots,\tilde{p}_{n_1+n_2},\dot{\tilde{p}}_{n_1+n_2}]^T$ . The unforced system $x^{(n_2)}=\mathbf{f}_{n_2}(x^{(n_2)}, 0)$ is given by
\begin{align*}
x^{(n_2)}=\mathbf{f}_{n_2}(x^{(n_2)}, 0), \quad \Rightarrow \quad
	\begin{cases}	
		\ddot{\tilde{p}}_{n_1+1}&=-f(\tilde{p}_{n_1+1})-g(\dot{\tilde{p}}_{n_1+1}),\\
				        &\vdots\\
		\ddot{\tilde{p}}_{n_1+n_2}&=-f(\tilde{p}_{n_1+n_2})-g(\dot{\tilde{p}}_{n_1+n_2}).
\end{cases}
\end{align*}
Again, following the same argument as  proving $x^{(n_1)}=\mathbf{f}_{n_1}(x^{(n_1)})$ is GES, we have that the origin of the unforced system $x^{(n_2)}=\mathbf{f}_{n_2}(x^{(n_2)}, 0)$ is GES. In addition, the functions $f, g$ are smooth enough functions, we obtain from~\cite[Lemma 4.6]{khalil} that the forced system $x^{(n_2)}=\mathbf{f}_{n_2}(x^{(n_2)}, x^{(n_1)})$ is ISS with $x^{(n_1)}$ as its input. We now invoke~\cite[Lemma 4.7]{khalil},  the origin of the cascade system $x^{(n_1+n_2)}=\mathbf{f}_{n_1+n_2}(x^{(n_1+n_2)})$ given in~\eqref{eq:cas} is globally asymptotically stable. By a straightforward induction method, we  prove that the origin $x=0$ of the whole formation $\dot{x}=\mathbf{f}(x)$ ($x= [\tilde{p}_1, \dot{\tilde{p}}_1,\cdots, \tilde{p}_N, \dot{\tilde{p}}_N]^T$) is globally asymptotically stable. \frQED
\end{proof-theorem}

\begin{proof-theorem}{\ref{thm:stability_bidi}}
For the vehicular formation with connected, undirected information graph,  we consider the following Lyapunov function candidate, which is inspired by the one used in~\cite{munz2011robust}:
\begin{align*}
	V(x)=\sum_{i=0}^{N} \sum_{j\in \scr{N}_i}\int_0^{\tilde{p}_i-\tilde{p}_{j}} f(z)  dz+  \sum_{i=0}^{N} \dot{\tilde{p}}_i^2,
\end{align*}
where $x= [\tilde{p}_1, \dot{\tilde{p}}_1,\cdots, \tilde{p}_N, \dot{\tilde{p}}_N]^T$. The derivative of $V$ along the trajectory of~\eqref{eq:coupled_ODE} with $w_i=0$ is 
\begin{align*}
\dot{V}=&\sum_{i=0}^{N} \sum_{j\in \scr{N}_i}  f(\tilde{p}_i-\tilde{p}_{j})  (\dot{\tilde{p}}_i-\dot{\tilde{p}}_{j})+2\sum_{i=0}^{N} \dot{\tilde{p}}_i  \ddot{\tilde{p}}_i \\ 
	=&\sum_{i=0}^{N} \sum_{j\in \scr{N}_i}  f(\tilde{p}_i-\tilde{p}_{j})  (\dot{\tilde{p}}_i-\dot{\tilde{p}}_{j})+2\sum_{i=0}^{N} \dot{\tilde{p}}_i  \Big(- \sum_{j \in \scr{N}_i} \Big (f(\tilde{p}_{i}-\tilde{p}_{j})+g(\dot{\tilde{p}}_{i}-\dot{\tilde{p}}_{j})\Big) \Big )\\
	= &- \sum_{i=0}^{N} \sum_{j\in \scr{N}_i}  (\dot{\tilde{p}}_i-\dot{\tilde{p}}_{j}) g(\dot{\tilde{p}}_i-\dot{\tilde{p}}_{j}) \leq 0. 
\end{align*}
If $\dot{V}=0$, then we have $\dot{\tilde{p}}_i=0$ for all $i$, since $g(z)$ satisfies $zg(z)>0, \forall x \neq 0$  and $\dot{\tilde{p}}_0=0$ by definition. Asymptotic stability now follows from LaSalle's Invariance Principle. In addition,  we have $V(x) \to \infty$ as $\|x\| \to \infty$, i.e. $V(x)$ is unbounded in each direction of the Euclidean coordinate of $x$. Therefore, the Lyapunov function $V$ is radially unbounded, and we get global asymptotic stability. \frQED
\end{proof-theorem}

\subsection{Numerical comparison between linear and nonlinear controllers for transient decay}\label{sec:conv_rate}
Throughout this section, we consider the following specific linear and nonlinear controllers. The control gain functions $f(z)$ and $g(z)$ used in controllers~\eqref{eq:u-pred} and \eqref{eq:u-bidi} are given by
\begin{align}\label{eq:control_function}
	\text{Linear: } &f(z)=k_0z,\ g(z) =b_0z,  \notag \\
	\text{Non-linear: } &f(z)=B_1 \tanh(\gamma_1 z),\ g(z)=B_2\tanh(\gamma_2 z),    
\end{align}
where $k_0=1,b_0=0.5, B_1=5, \gamma_1=0.2, B_2=5, \gamma_2=0.1$.  The  parameters have been chosen in such a way that the slopes of $f(z)$ of $g(z)$ near the origin are equal to $k_0$ and $b_0$, respectively. This is done to make the linear and non-linear cases comparable to some extent. Note that these $f(z)$ and $g(z)$ do not satisfy the \emph{global} sector conditions assumed in  Theorem~\ref{thm:stability_pred}, but only satisfy the sector conditions \emph{locally}. However, the region in which they satisfy the sector condition can be made arbitrarily large by choosing sufficiently small $\varepsilon_1$ and  $\varepsilon_2$. 

\begin{figure}
	  \psfrag{time}{\scriptsize time ($t$)}
	  \psfrag{position}{\small $\tilde{p}_N$}
	  \psfrag{linear}{\scriptsize Linear}
	  \psfrag{nonlinear}{\scriptsize Nonlinear}
\centering
\subfigure[Predecessor-following]{\includegraphics[scale = 0.35]{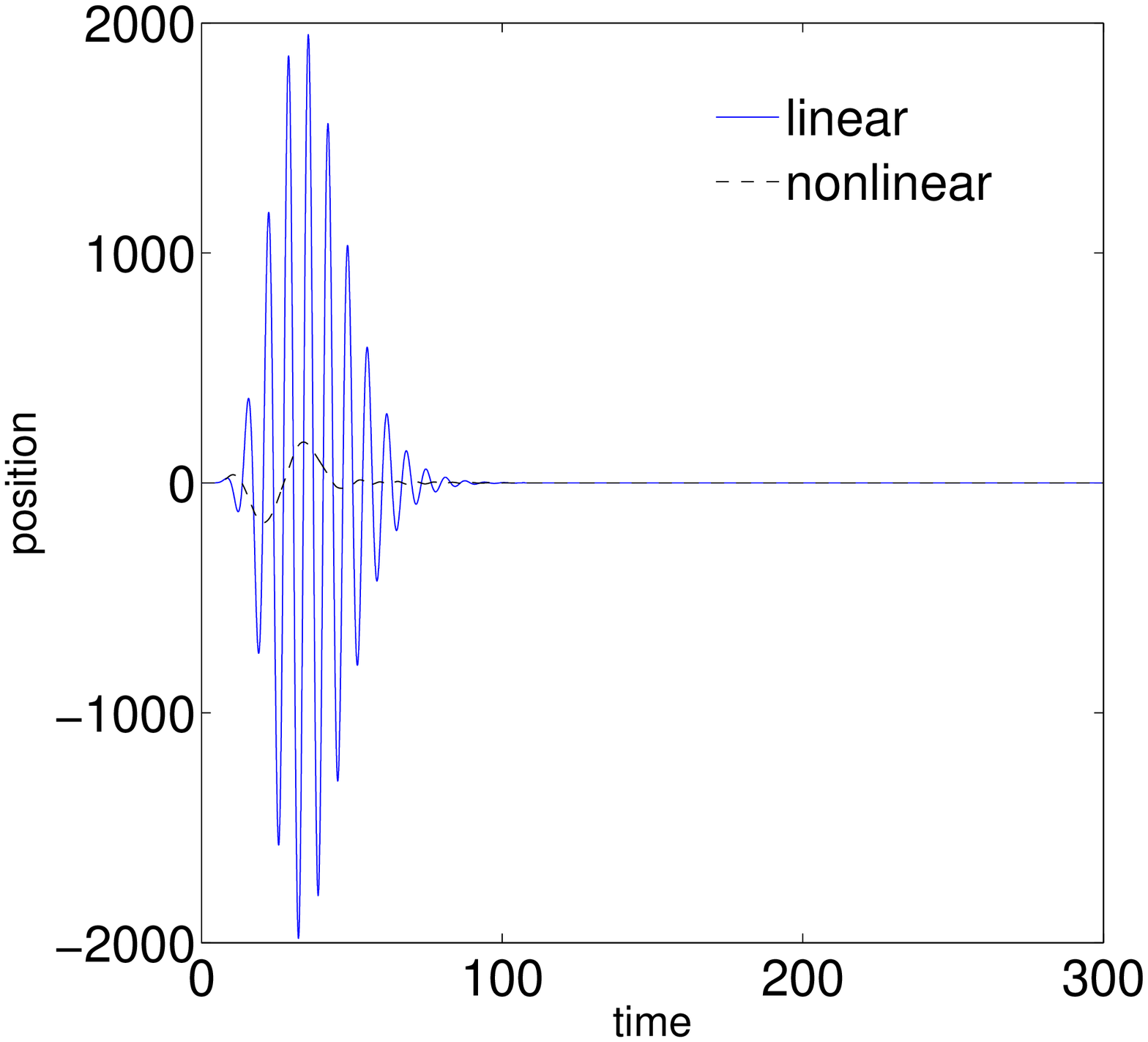}}  
\subfigure[Symmetric bidirectional]{\includegraphics[scale = 0.35]{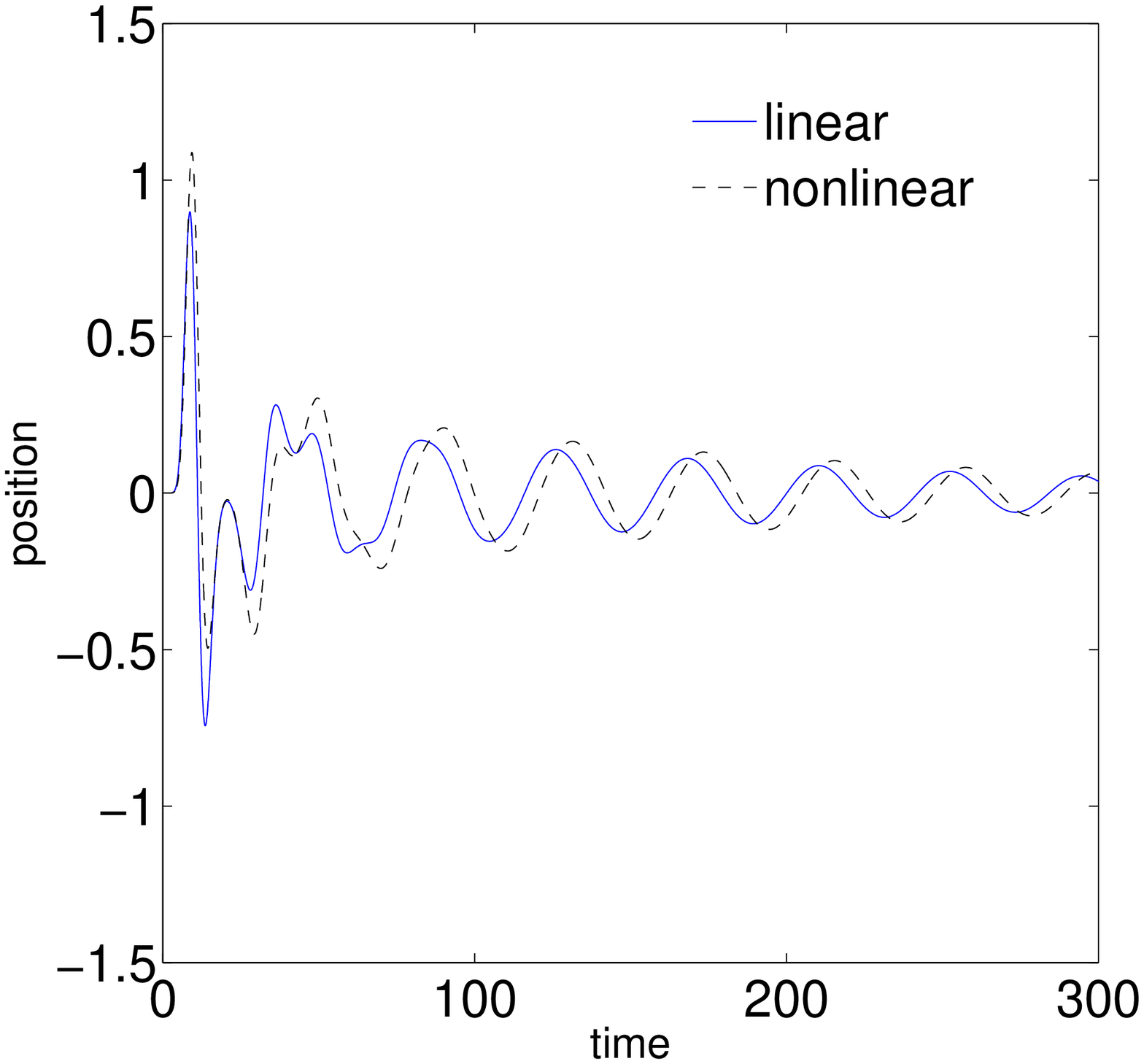}} 
\caption{Comparison of transients of the position tracking error of the last vehicle for a platoon of $N=10$ vehicles between linear and nonlinear controller. The initial condition of the first vehicle used is $x_0=10$.}
\label{fig:transient}
\end{figure}

In this section, we compare the convergence rate and transient performance between linear and nonlinear controllers through numerical simulations.  Figure~\ref{fig:transient} (a) depicts the transients of the platoon with linear and nonlinear controllers for predecessor-following architecture. The algebraic growth for linear controller which is predicted by Theorem~\ref{thm:eig_relation} is observed.  We also see that the nonlinear controller has much smaller peak error than the linear controller. The transients of the symmetric bidirectional architecture are shown in Figure~\ref{fig:transient} (b). We see that (i) the performance of the non-linear case is similar to that of the linear controller, and (ii) the peak value of the error is much smaller compared to that in the predecessor-following architecture.

\begin{figure}
	  \psfrag{time}{\scriptsize time ($t$)}
	  \psfrag{position}{\small $\tilde{p}_N$}
	  \psfrag{Linear predecessor-following}{\scriptsize Linear predecessor-following}
          \psfrag{Nonlinear predecessor-following}{\scriptsize Nonlinear predecessor-following}
	  \psfrag{Linear symmetric bidirectional}{\scriptsize Linear symmetric bidirectional}
 	  \psfrag{Nonlinear symmetric bidirectional}{\scriptsize Nonlinear symmetric bidirectional}
\centering
\includegraphics[scale = 0.4]{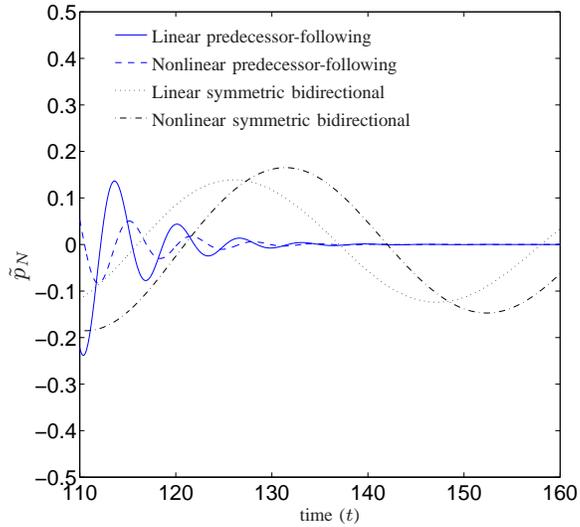}
\caption{Comparison of transients of the position tracking error of the last vehicle for a platoon of $N=10$ vehicles between predecessor-following and symmetric bidirectional architectures. The initial condition of the first vehicle used is $x_0=10$.}
\label{fig:transient2}
\end{figure}

Moreover, we see from Figure~\ref{fig:transient2} that the convergence rates of the linear and non-linear controllers in the predecessor-following architecture are similar. In addition, the error in the predecessor-following architecture  is smaller than in the case of symmetric bidirectional architecture for \emph{large $t$}, irrespective of whether the control is linear or non-linear. This is due to the fact that in  the predecessor-following architecture, the stability margin is independent of the number of vehicles $N$ in the platoon, while it's very small when $N$ is large for symmetric bidirectional architecture. 

\begin{figure}
	\psfrag{V}{\scriptsize $\hat{E}$}
	  \psfrag{N}{\scriptsize $N$}
	  \psfrag{linear}{\scriptsize Linear}
	  \psfrag{nonlinear}{\scriptsize Nonlinear}
\centering
\subfigure[Predecessor-following]{\includegraphics[scale = 0.34]{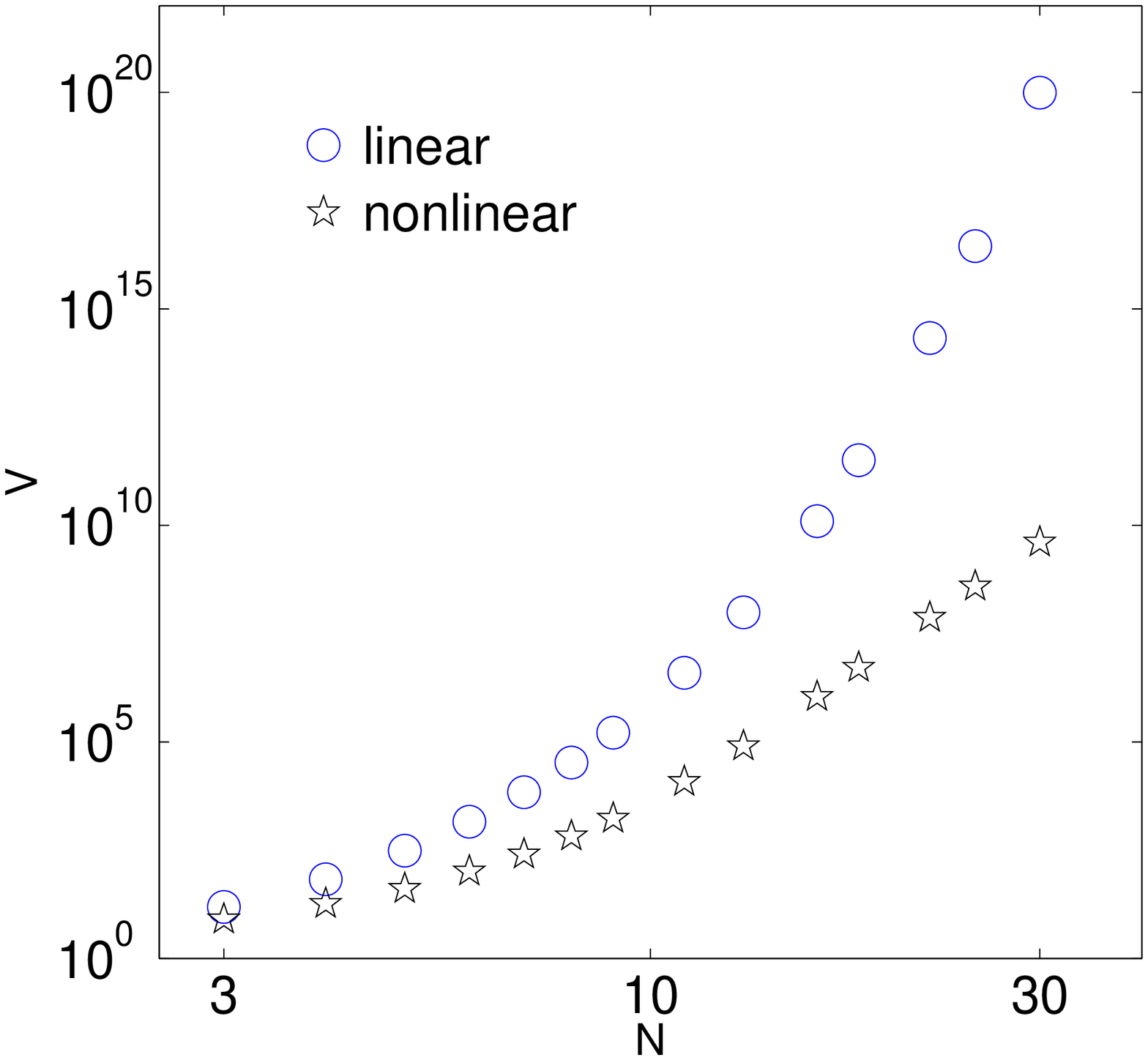}}
\subfigure[Symmetric bidirectional]{\includegraphics[scale = 0.34]{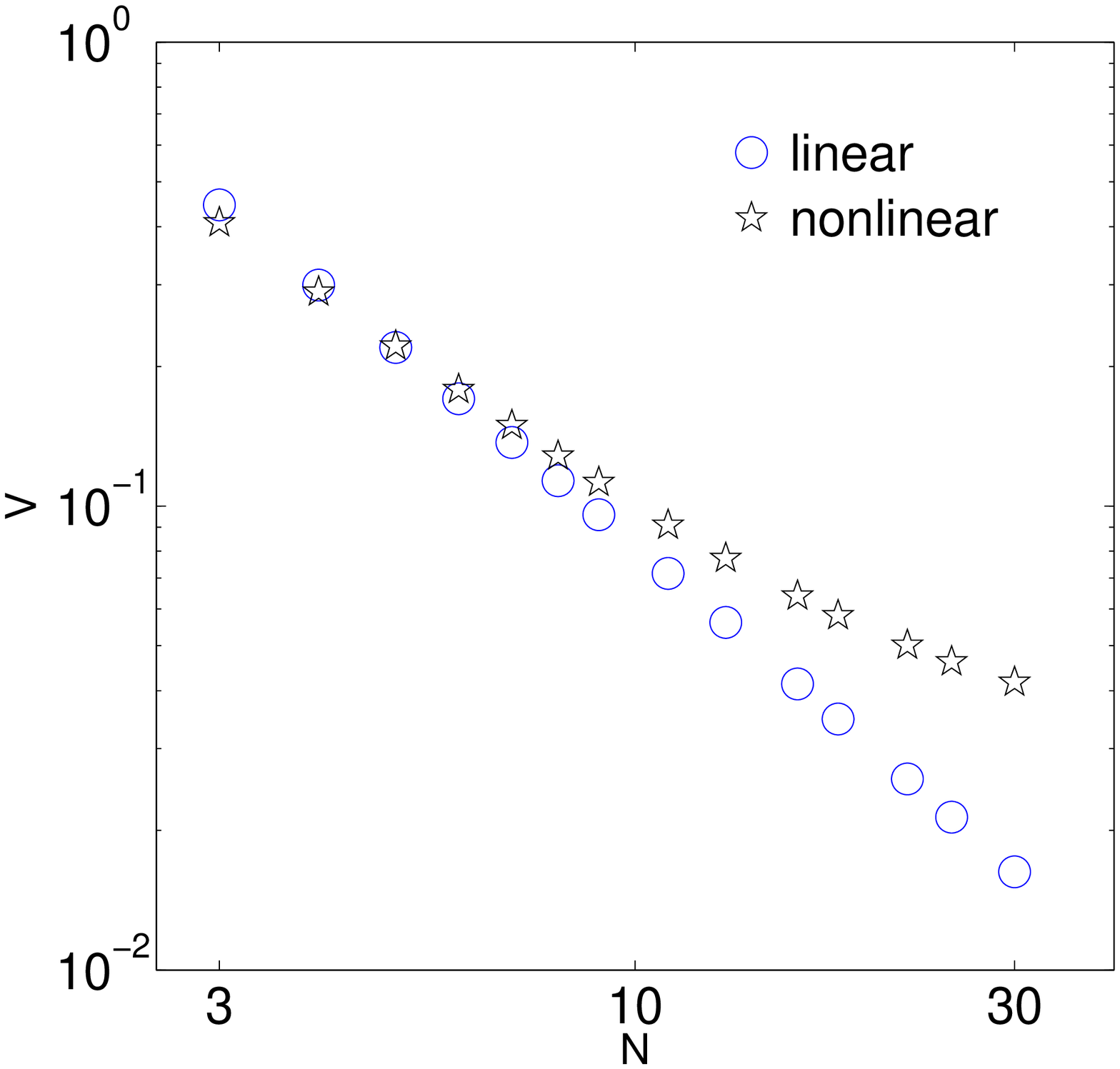}}

\caption{Comparison of $E$ between linear and nonlinear controllers as a function of $N$. The measure $E$ is estimated by numerically evaluating the integral in~\eqref{eq:energy} for $T=10^4$ s. The initial condition of the first vehicle used is $x_0=10$.}
\label{fig:energy}
\end{figure}

\begin{figure}[h]
	\psfrag{V}{\scriptsize $\hat{E}$}
	  \psfrag{N}{\scriptsize $x_0$}
	  \psfrag{linear}{\scriptsize Linear}
	  \psfrag{nonlinear}{\scriptsize Nonlinear}
\centering
\subfigure[Predecessor-following]{\includegraphics[scale = 0.34]{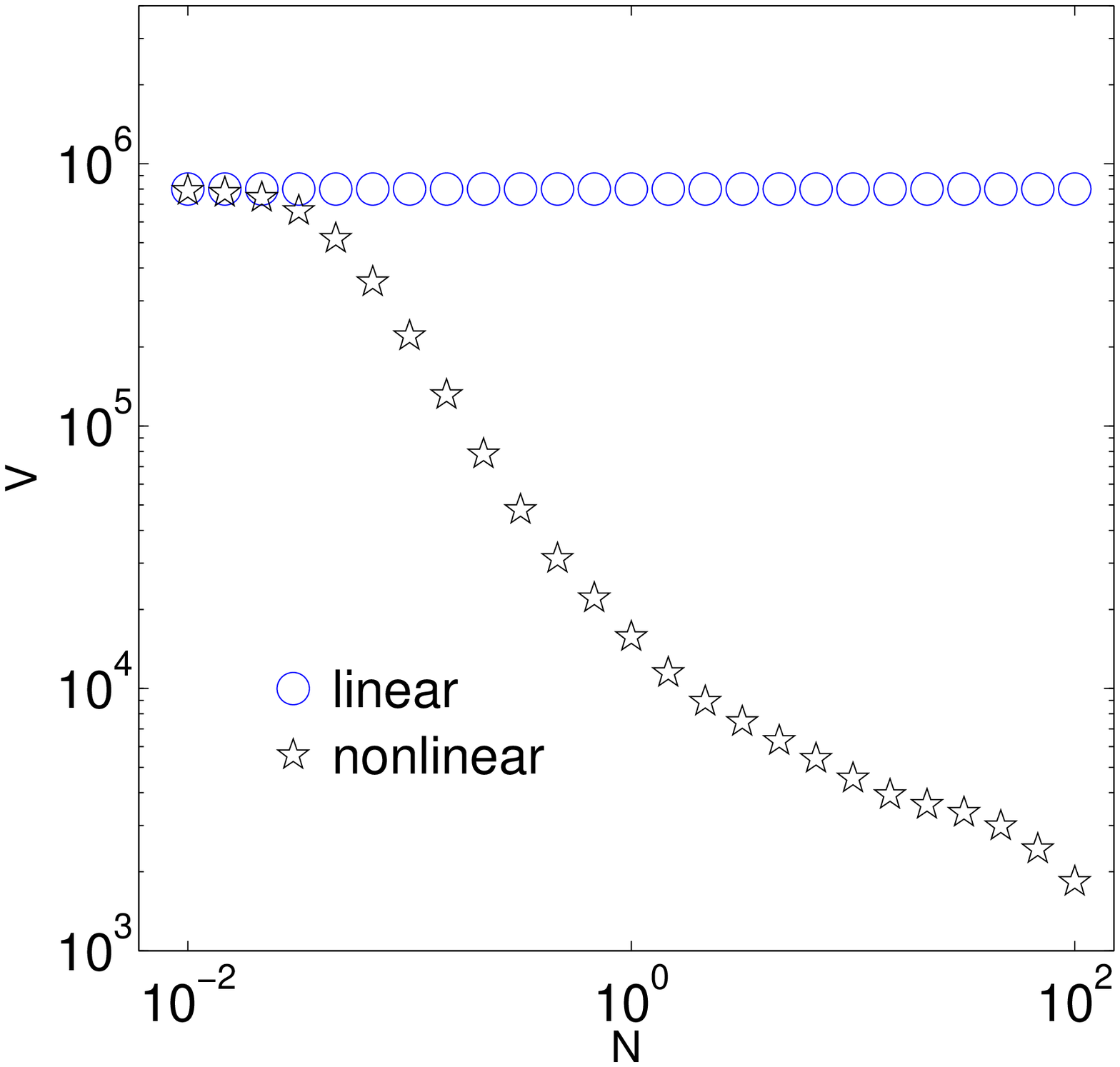}}
\subfigure[Symmetric bidirectional]{\includegraphics[scale = 0.34]{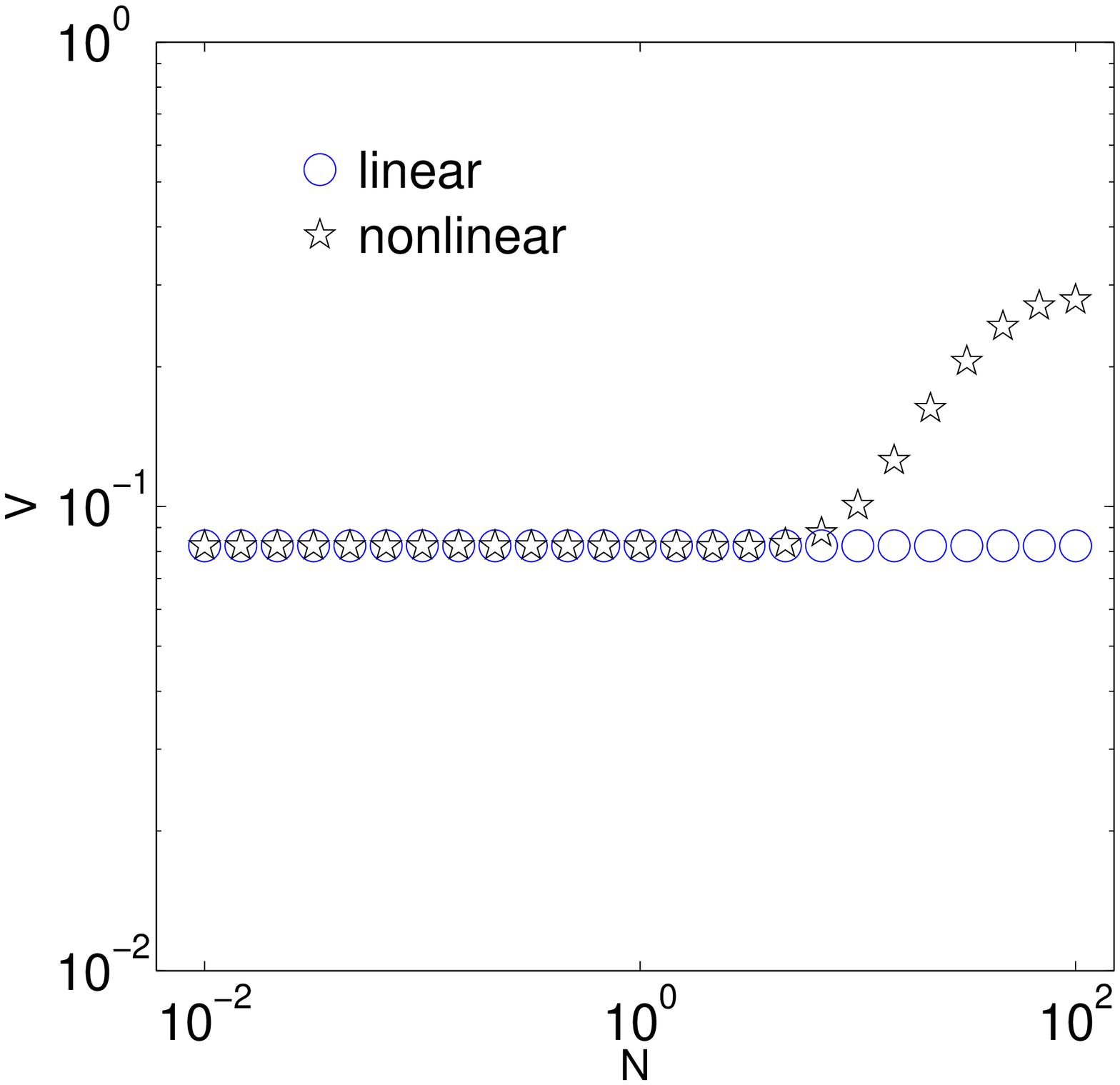}}

\caption{Comparison of $\hat{E}$ between linear and nonlinear controllers for a platoon of $N=10$ vehicles with different initial conditions $x_0$. The measure $\hat{E}$ is estimated by numerically evaluating the integral in~\eqref{eq:energy} for $T=10^4$ s.}
\label{fig:energy_amplitude}
\end{figure}

Figure~\ref{fig:energy} and Figure~\ref{fig:energy_amplitude} show the estimate of  energy measure $\hat{E}$ for $T=10^4$ seconds (defined in~\eqref{eq:energy}) as a function of $N$ and $x_0$ respectively. We see that (i) the  energy in  the predecessor-following architecture  has a much worse scaling trend with $N$ or $x_0$ than that in symmetric bidirectional architecture, no matter the controller is linear or nonlinear, (ii) nonlinear controller performs better than linear controller in the predecessor-following architecture, whereas it performs similarly or worse in the symmetric bidirectional architecture. 

In summary, the predecessor-following architecture has a faster convergence rate than that of symmetric bidirectional architecture, no matter the controller is linear or nonlinear. However, due to the high algebraic growth of initial errors, the predecessor-following architecture has a worse scaling trend of  energy, comparing to the symmetric bidirectional architecture.  Moreover, if only one architecture is considered, the nonlinear controller in general has a better (respectively, worse) transient performance than that of linear controller in the predecessor-following (respectively, symmetric bidirectional) architecture.   

\section{Sensitivity to external disturbances}\label{sec:h8norm}
In this section, we study the sensitivity of the vehicular formation to external disturbances. Specifically, we examine appropriate gains from disturbances acting on all vehicles $w\in \R^N$ to the position tracking errors $\tilde{p} \in \R^N$. We define the \emph{amplification factor} $AF$ as the $L_2$ gain from the vector of disturbances $w(t)=[w_1(t),\cdots, w_N(t)]$ to position tracking error vector $\tilde{p}(t)=[\tilde{p}_1(t),\cdots,\tilde{p}_N(t)]$:
\begin{align}\label{eq:sinusoidal}
	AF^{linear \text{ or } nonlnear} = \sup \frac{\|\tilde{p}\|_{\scr{L}_2}}{\|w\|_{\scr{L}_2}}. 
\end{align}
In this paper, the $\L_2$ norm is well-defined in the extended space $\L^e_2=\{u|u_\tau \in \L_2, \forall \ \tau \in[0,\infty) \}$, where $u_\tau(t) = (i)\ u(t), \text{ if } 0\leq t\leq \tau; (ii)\ 0, \text{ if }\ t>\tau.$ See~\cite[Chapter 5]{khalil}. With a little abuse of notation, we suppress the subscript and write $\L_2=\L^e_2$.

In the linear case the amplification factor is equivalent to the $H_\infty$ norm of the transfer function $G(s)$ from $w$ to $\tilde{p}$,
\begin{align*}
	AF^{linear} & = \max_{\omega}  \sigma_{\max} (G(j\omega))=\sigma_{\max} (G(j\omega_p)),  
\end{align*}
where we have assumed the maximum is achieved, and $ \omega_p \eqdef \arg\max_{\omega} \sigma_{\max} (G(j\omega))$, $\sigma_{\max}$ denotes the maximum singular value. In the non-linear case, evaluating  $AF^{nonlinear} $ is intractable, so we use following conservative estimate:
\begin{align}\label{eq:sinusoidal1}
	\hat{AF}^{nonlinear}& \eqdef \frac{\|\tilde{p}\|_{\scr{L}_2(\tau)}}{\|w\|_{\scr{L}_2(\tau)}}, 
\end{align}
where $\|u(t)\|_{\L_2(\tau)}=\sqrt{\int_0^\tau \sum_i u_i^2(t) dt}$, $\tau$ is large enough and $w  = [a_1 \sin(\omega_p t+\theta_1), \cdots, a_N \sin(\omega_p t+\theta_N)]$. The parameters $a=[a_1,\cdots,a_N]$ and $\theta=[\theta_1,\cdots,\theta_N]$ are those that yield the $\L_2$ gain in the linear case. The choice of these parameter will be given later. Note that the estimates for the non-linear case are lower bounds: $\hat{AF}^{nonlinear} \leq AF^{nonlinear}$. In the numerical simulations, the quantity $AF^{linear}$ for linear controller are also estimated for large $\tau$ to compare with analytical results, which will be denoted by $\hat{AF}^{linear}$, as defined in~\eqref{eq:sinusoidal1}.

We also examine the effect of random disturbances. Specifically, let $w(t)$ in the closed-loop platoon dynamics~\eqref{eq:ss-nl} be a vector of white noise with autocorrelation matrix $\sigma_0 I$, where $\sigma_0$ is a constant and $I$ is the identity matrix with appropriate dimension.  Similar to sinusoidal disturbances, we define the following metric
\begin{align}\label{eq:R-def1}
 R^{linear \text{ or } nonlnear} \eqdef \lim_{t \to \infty}\frac{\sqrt{E(\tilde{p}(t)^T\tilde{p}(t))}}{\sigma_0},
\end{align}
where $E(.)$ denotes the expected value and we have assumed the above limits exist. Notice  in the linear case, the above ratios are exactly the $H_2$ norms of the appropriate transfer functions from the white noise disturbances to the position tracking errors. The steady-state covariance matrix of the state $\tilde{p}(t)$ of the system~\eqref{eq:ss} that is driven by a white noise process $w(t)$ is given by solution $P$ of the following Lyapunov equation~\cite[Chapter 4]{simon_estimation}: 
\begin{align*}
	AP+PA^T=-Q,
\end{align*}
where $Q= E[B w w^T B^T]$, and $B$ is the input matrix. Since $A$ is Hurwitz, it guarantees the limit in~\eqref{eq:R-def1} exists~\cite{simon_estimation}. The steady-state expectations $E(\tilde{p}(t)^T\tilde{p}(t))$ can be obtained  by   summing the odd diagonal entries of $P$, which yields 
\begin{align} \label{eq:lya_equation} 
	R^{linear}=\frac{\sqrt{\sum_{i=1}^N P(2i-1,2i-1)}}{\sigma_0}.
\end{align}
For the non-linear controllers as well as linear controllers,  we use the following estimate of the ratio defined in~\eqref{eq:R-def1},
\begin{align}\label{eq:R-def}
	 \hat{R}^{linear \text{ or } nonlnear}\eqdef \frac{\sqrt{E(\tilde{p}(T)^T\tilde{p}(T))}}{\sigma_0},
\end{align}
where $T$ is sufficiently large. Monte-Carlo simulations are used to estimate the ratios. For example, to estimate the  ratio $\hat{R}^{nonlinear}$ for the 1-D vehicular platoon with predecessor-following architecture, the noise-driven system is converted into a standard stochastic differential equation (SDE) form
\begin{align}\label{eq:SDE}
d \tilde{p}_i& =\dot{\tilde{p}}_i dt, \quad d \dot{\tilde{p}}_i = -f(\tilde{p}_{i}-\tilde{p}_{i-1}) dt -g (\dot{\tilde{p}}_{i}-\dot{\tilde{p}}_{i-1}) dt+\sigma_0 dW_i(t),
\end{align}
where $W(t)=[W_1(t),\cdots,W_N(t)]$ is a standard $N$-dimensional Wiener process.  Sample paths of the states are computed by using Euler-Maruyama Method to numerically integrate the SDE~\eqref{eq:SDE}~\cite{SDE}. The ratio $\hat{R}^{nonlnear}$ is now estimated by performing appropriate averaging over a large number of simulations, after letting each simulation proceed sufficiently long to allow transients to  die out.

\subsection{Sensitivity to disturbance with linear control}
As stated earlier, analytical results on  the sensitivity to disturbance is possible only for the linear case. The first result is on the sensitivity of the vehicular formation with directed tree graph.
\begin{thm}\label{thm:pre}
Consider a vehicular formation with directed tree information graph who has a spanning tree. Let $k$ be the number of generations. With linear controller $f(x)=k_0x$ and $g(x)=b_0x$ in~\eqref{eq:u}, the amplification factor $AF^{linear}$ satisfies 
\begin{align*}
	\beta_1 \alpha^{k-1}  \leq	AF^{linear} \leq \frac{\beta_2 (\alpha^k-1)}{\alpha-1},
\end{align*}
where $\alpha=|T(j\omega_T)|>1$, $\beta_1=|S(j\omega_T)|$ and $\beta_2=|S(j\omega_S)|$, in which
\begin{align*}
	T(s)=\frac{b_0s+k_0}{s^2+b_0s+k_0}, \quad S(s)=\frac{1}{s^2+b_0s+k_0},
\end{align*}
and $\omega_T$ and $\omega_S$ are the peak frequencies of $T(s)$ and $S(s)$ respectively.\frqed
\end{thm}
Notice that in the above theorem, if the directed tree has bounded degree $d$, then we have $k\to \infty$ as $N \to \infty$. The amplification factor is unbounded for increasing $N$.

\begin{proof-theorem}{\ref{thm:pre}}
	The proof is inspired by the proof of Lemma $1$ in~\cite{Seiler_disturb_TAC:04}. With linear controller $f(x)=k_0x$ and $g(x)=b_0x$, taking Laplace transform of the coupled-ODE model~\eqref{eq:coupled_ODE} and assuming zero initial conditions, we obtain the transfer function from the disturbances $w = [w_1,\dots,w_N]^T$ to position errors $\tilde{p} =[ \tilde{p}_1,\dots,\tilde{p}_N]^T$
\begin{align*}
	G(s)= (s^2 I_N+(b_0s+k_0)L_g^{(d)})^{-1},
\end{align*}
where $I_N$ is the $N \times N$ identity matrix and $L_g^{(d)}$ is the grounded graph Laplacian of the directed tree. It follows from $\|G\|_{\text{max}} \leq \|G\|_2$ of~\cite[Chapter 10]{matrix_cookbook} that the $H_\infty$ norm of the transfer function $G$ has the following relation
\begin{align}\label{eq:lower}
	AF^{linear}=	\|G\|_{H_\infty} \geq \|G(j\omega_T)\|_2 \geq \|G(j\omega_T)\|_\text{max}=\max_{i,j} |G_{ij}(j\omega_T)|.
\end{align}

We now look at a particular entry of $G(s)$. Since there is $k$ generations in the directed tree, let $(0,p_1),(p_1,p_2),\cdots,(p_{k-1},p_k)$ be a longest path starting from the root.  Its closed-loop dynamics can be written as
\begin{align*}
	\ddot{\tilde{p}}_{p_i} &= - k_0 (\tilde{p}_{p_i}-\tilde{p}_{{p_{i-1}}})-b_0 (\dot{\tilde{p}}_{p_i}-\dot{\tilde{p}}_{p_{i-1}})+w_{p_i}, \quad i \in \{1,\cdots, k\}.
\end{align*}
Taking Laplace transform of the above coupled-ODEs and assuming zero initial conditions, we have the transfer function from the disturbance $w_{p_1}$ on the first vehicle to position tracking error $\tilde{p}_{p_k}$ of the last vehicle 
\begin{align}\label{eq:particular}
	G_{w_1 p_k}(s)=\frac{P_{p_k}}{W_{p_1}}=\frac{(b_0s+k_0)^{k-1}}{(s^2+b_0s+k_0)^k}=S(s) T(s)^{k-1},
\end{align}
where $P_{p_k}$ and $W_{p_1}$ are the Laplace transforms of $\tilde{p}_{p_k}$ and $w_{p_1}$ respectively. In addition, $|G_{w_1 p_k}(j\omega_T)| \leq \max_{i,j} |G_{ij}(j\omega_T)|$, we obtain from~\eqref{eq:lower} a lower bound for $AF^{linear}$
\begin{align*}
	AF^{linear} \geq |G_{w_1 p_k}(j\omega_T)|= \beta_1 \alpha^{k-1},
\end{align*}
where $\alpha=|T(j\omega_T)|, \beta_1=|S(j\omega_T)|$. In particular,  straightforward algebra shows that $\alpha>1$.

In addition, for any matrix $A$, its spectral radius satisfies $\rho(A) \leq \|A\|_1$, we have
\begin{align}\label{eq:upper}
	\|G\|_{H_\infty}&=\sup_\omega \|G(j\omega)\|_2 = \sup_\omega \sqrt{ \rho(G^*(j\omega)G(j\omega))} \leq \sup_\omega \sqrt{\|G^*(j\omega)G(j\omega)\|_1} \notag\\ &\leq \sup_\omega \sqrt{\|G^*(j\omega)\|_1 \|G(j\omega)\|_1 } = \sup_\omega \sqrt{\|G(j\omega)\|_\infty \|G(j\omega)\|_1 } \notag \\&\leq \sup_\omega \sqrt{\|G(j\omega)\|_\infty} \sup_\omega \sqrt{\|G(j\omega)\|_1 }.
\end{align}
As the particular entry  $G_{w_1 p_k}(s)$ given in~\eqref{eq:particular}, each entry of $G(s)$ is either $0$ or $S(s) T(s)^{j-1}$ for some $1\leq j\leq k$. The absolute row sum of $G(s)$ is $\sum_j |G_{ij}(j\omega)|$, where  $G_{ij}(j\omega)$ is the transfer function from disturbance $w_j$ to $\tilde{p}_i$. Because of the special coupling structure of the directed tree,  $G_{ij}(j\omega)=0$ if $j$  is not a predecessor of $i$ and $G_{ij}(j\omega)=S(s) T(s)^{l-1}$ if $j$  is the last $(l-1)$-th predecessor of $i$, i.e. if $j=i$, then $l=1$; if $j$ is the parent of $i$, then $l=2$ and so forth. Since there is $k$ generation in the directed tree, the maximum absolute row sum is given by 
\begin{align*}
	\|G(j\omega)\|_\infty=  \max_i \sum_j |G_{ij}(j\omega)|=\sum_{l=1}^k |S(j\omega)| |T(j\omega)|^{l-1}.
\end{align*}
In addition, we have $\sup_\omega |S(j\omega)|| T(\omega)^{j-1}| \leq \sup_\omega|S(j\omega)| \sup_\omega |T(\omega)^{j-1}| = \beta_2 \alpha^{j-1}$, where $\beta_2=|S(j\omega_S)|$. Thus we have
\begin{align}\label{eq:h8}
	\sup_\omega \|G(j\omega)\|_\infty = \sup_\omega \sum_{l=1}^k |S(j\omega)| |T(j\omega)|^{l-1} \leq  \sum_{l=1}^k \beta_2 \alpha^{l-1}=\beta_2\frac{\alpha ^k-1}{\alpha-1}.
\end{align}
Following the same argument as before, we obtain that 
\begin{align}\label{eq:h1}
	\sup_\omega \|G(j\omega)\|_1 \leq \beta_2\frac{\alpha ^k-1}{\alpha-1}.
\end{align}
Substituting~\eqref{eq:h8} and~\eqref{eq:h1} into~\eqref{eq:upper}, we obtain a upper bound for $AF^{linear}$
\begin{align*}
	\|G\|_{H_\infty} \leq \beta_2\frac{\alpha ^k-1}{\alpha-1}. \tag*{\frQED}
\end{align*}
\end{proof-theorem}

\begin{corollary}\label{cor:pre}
	Consider an $N$-vehicle platoon with predecessor-following architecture. With linear controller $f(x)=k_0x$ and $g(x)=b_0x$ in~\eqref{eq:u-pred}, the  amplification factor $AF^{linear}$ satisfies
\begin{align*}
	\beta_1 \alpha^{N-1}  \leq	AF^{linear} \leq \frac{\beta_2 (\alpha^N-1)}{\alpha-1}.
\end{align*}
Furthermore, when $N \gg 1$, 
\begin{align}\label{eq:limit}
	AF^{linear}  & \approx \beta_1\sqrt{\frac{ (\alpha^{2N}-1)}{\alpha^2-1}}, & \omega_p\approx &\frac{\sqrt{\sqrt{k_0^4+2k_0^3b_0^2}-k_0^2}}{b_0}.
\end{align}
Moreover, a sufficient condition for a disturbance $w=[w_1,\cdots, w_N]=[a_1\sin(\omega t+ \theta_1), \cdots, a_N\sin(\omega t+ \theta_N)]$ to yield the amplification factor is $a=[a_1,\cdots, a_N]=[a_1, 0, \cdots, 0]$, where $a_1$ is an arbitrary constant and $\omega=\omega_p$, $\theta=[\theta_1, \cdots, \theta_N]=0$.\frqed
\end{corollary}

\begin{proof-corollary}{\ref{cor:pre}}
The inequality of $AF^{linear}$ follows immediately by noting that in the predecessor-following architecture, the information graph is a line graph who has $N$ generations. 
	
To prove the asymptotic results,  the $H_\infty$ norm of a system  can be interpreted in a sinusoidal, steady-state sense as follows (see~\cite{h8norm}). For any frequency $\omega$, any vector of amplitudes $a=[a_1,\cdots,a_N]$  and any vector of phases $\theta=[\theta_1,\cdots,\theta_N]$, the input vector
\begin{align*}
w=[w_1, \cdots, w_N] = [a_1\sin(\omega t+\theta_1),\cdots,a_N\sin(\omega t+\theta_N)]
\end{align*}
yields the  steady-state response $e$ of the form
\begin{align*}
	e=[\tilde{p}_1, \cdots, \tilde{p}_N] = [b_1\sin(\omega t+\psi_1),\cdots,b_N\sin(\omega t+\psi_N)].
\end{align*}
The $H_\infty$ norm of $G(j\omega)$ can be defined as
\begin{align}\label{eq:norm}
	\|G(j\omega)\|_{H_\infty}=\sup_{\omega\in \R^+, a,\theta \in \R^N} \frac{\|\bar{x}\|_{\scr{L}_2}}{\|w\|_{\scr{L}_2}}.
\end{align}
Because of the special coupling structure of the predecessor-following architecture, the disturbance acting on a vehicle only effect its descendants. In addition, the supremum of the norm of  the complementary sensitivity function $\sup_\omega |T(s)| >1$, therefore to get the worst amplification factor, the disturbance should all concentrate on the first vehicle, that is the other vehicles are undisturbed ($w_i=0$ for $i\in \{2,\cdots, N\}$). In addition,  the steady state response correspond to a sinusoidal disturbance on the first vehicle $w_1=a_1\sin(\omega t)$ are given by $\tilde{p}_i(t)=a_1 |T(j\omega)^{i-1}S(j\omega)| \sin(\omega t+\psi_i)$.  It follows~\eqref{eq:norm}, the $H_{\infty}$ norm of the system is given by
\begin{align*}
	\|G(j\omega)\|_{H_\infty} &= \sup_\omega \frac{\|[a_1 |S(j\omega)| \sin(\omega t+\psi_1), \cdots, a_1|T(j\omega)^{i-1}S(j\omega)| \sin(\omega t+\psi_N)]\|_{\L_2}}{\|a_1\sin(\omega t)\|_{\scr{L}_2}}\\ &= \sup_\omega \sqrt{\sum_{\ell=1}^N|T(j\omega)^{\ell-1}S(j\omega)|^2} = \sup_\omega \sqrt{|S(j\omega)|^2 \frac{|T(j\omega)|^{2N}-1}{|T(j\omega)|^2-1}}.
\end{align*}
Since $\alpha=\sup_\omega |T(j\omega)|>1$, in the limit of $N \to \infty$, the supremum of the above equation is determined by $\sup_\omega |T(j\omega)|^{2N}$, thus we have 
\begin{align*}
	\|G(j\omega)\|_{H_\infty}=\sqrt{|S(j\omega_T)|^2 \frac{|T(j\omega_T)|^{2N}-1}{|T(j\omega_T)|^2-1}}=\beta_1 \sqrt{\frac{\alpha^{2N}-1}{\alpha^2-1}},
\end{align*}
Following straightforward algebra, we have $\omega_p=\omega_T=\frac{\sqrt{\sqrt{k_0^4+2k_0^3b_0^2}-k_0^2}}{b_0}$, which completes the proof.  \frQED
\end{proof-corollary}

The next theorem is the corresponding result for the vehicular formation with undirected information graph.
\begin{theorem}\label{thm:bidi}
Consider a vehicular formation with connected, undirected information graph. With linear controller $f(x)=k_0x$ and $g(x)=b_0x$ in~\eqref{eq:u}, the amplification factor  and its peak frequency satisfy
\begin{align*}
AF^{linear} =\begin{cases} 
	\frac{2}{\lambda_1^{3/2} b_0\sqrt{4 k_0 -\lambda_1 b_0^2}},   & \mbox{if } \lambda_1 \leq 2k_0/b_0^2, \\
	\frac{1}{\lambda_1 k_0}, & \mbox{otherwise.} 
\end{cases}, \quad 
\omega_p=  
	\begin{cases} 
		\frac{\sqrt{4\lambda_1 k_0-2\lambda_1^2 b_0^2}}{2},  & \mbox{if } \lambda_1 \leq 2k_0/b_0^2,\\
		0, & \mbox{otherwise.} 
\end{cases}
 \end{align*}
 where $\lambda_1$ is the smallest eigenvalue of the grounded graph Laplacian $L_g^{(u)}$. 
 
 Moreover, a sufficient condition for a disturbance $w=[w_1,\cdots, w_N]=[a_1\sin(\omega t+ \theta_1), \cdots, a_N\sin(\omega t+ \theta_N)]$ to yield the amplification factor is $a=[a_1,\cdots, a_N]=v_1$, where $v_1$ is the eigenvector of $L_g^{(u)}$ corresponding to the principal eigenvalue $\lambda_1$ and $\omega=\omega_p$ and $\theta=[\theta_1, \cdots, \theta_N]=0$.\frqed
\end{theorem}
\begin{remark}
	The above result is complimentary to Theorem $3.1$ of~\cite{SKY_SD_KRR_TAC:06} and Theorem $6$ of~\cite{SD_PP_IJES:10}. In~\cite{SKY_SD_KRR_TAC:06}, the authors considered undirected information graph whose maximum degree is a function $q(N+1)$, where $N+1$ is the number of vehicles (including the reference vehicle) in the formation. It follows from Lemma 3.2 of~\cite{SKY_SD_KRR_TAC:06} that $\lambda_1 \leq q(N+1)/{N}$. From Theorem~\ref{thm:bidi}, the amplification factor is $O((N/q(N+1))^{3/2})$, which is the same as the result of Theorem $3.1$ in~\cite{SKY_SD_KRR_TAC:06}. In~\cite{SD_PP_IJES:10}, the authors considered the undirected information whose $\lambda_1$ is $O(1/N)$, and derived an lower bound of $H_\infty$ norm by letting the peak frequency to be $0$. In the context of Theorem~\ref{thm:bidi}, it yields $AF^{linear}=\frac{1}{\lambda_1 k_0}$, which implies $AF^{linear}$ is at least $O(N)$. This coincides with the result of Theorem $6$ in~\cite{SD_PP_IJES:10}.\frqed
\end{remark}
\begin{proof-theorem}{\ref{thm:bidi}}
 It follows from straightforward algebra that the transfer function from the disturbance $w = [w_1,\dots,w_N]^T$ to position error $\tilde{p} =[ \tilde{p}_1,\dots,\tilde{p}_N]^T$ is given by
\begin{align*}
	G(s)= (s^2 I_N+(b_0s+k_0)L_g^{(u)})^{-1},
\end{align*}
where $L_g^{(u)}$ is the grounded graph Laplacian of the vehicular formation. Since $L_g^{(u)}$ is real and symmetric, there exists an orthogonal matrix $P$ such that  $L_g^{(u)}=P \Lambda P^T$, where $\Lambda=\diag\{\lambda_1, \cdots, \lambda_N\}$. The transfer function now becomes
\begin{align*}
	G(s)=(s^2 I+(b_0s+k_0)L)^{-1}
	    =P \begin{bmatrix}
		    G_1(s) & &   & \\
		 & \ddots &   &\\
		 &  & &  G_N(s)
	       \end{bmatrix} P^T ,
\end{align*}
where $G_\ell(s)=\frac{1}{s^2+b_0\lambda_\ell s+\lambda_\ell k_0}$. It can be shown using straightforward calculus that for each eigenvalue $\lambda_\ell$, the maximum amplitude  and its peak frequency of $G_\ell(s)$ are
\begin{align}
	A_\ell& \eqdef \max_{\omega} |G_\ell(j\omega)|=
\begin{cases} 
	\frac{2}{\lambda_\ell^{3/2} b_0\sqrt{4 k_0 -\lambda_\ell b_0^2}}, \label{eq:amplitude2}  & \mbox{if } \lambda_\ell \leq 2k_0/b_0^2, \\
	\frac{1}{\lambda_\ell k_0}, & \mbox{otherwise.} 
\end{cases}\\
	\omega_\ell& \eqdef \arg \max |G_\ell(j\omega)| = 
	\begin{cases} 
		\frac{\sqrt{4\lambda_\ell k_0-2\lambda_\ell^2 b_0^2}}{2}, \label{eq:frequency2}  & \mbox{if } \lambda_\ell \leq 2k_0/b_0^2,\\
		0, & \mbox{otherwise.} 
\end{cases}
\end{align}
It follows from straightforward algebra that $A_1 \geq A_2 \geq \cdots \geq A_N$. 

The $H_\infty$ norm of $G(s)$ is now given by
\begin{align*}
	\|G\|_{H_\infty} =& \sup_\omega \|G(j\omega)\|_2=  \sup_\omega \sqrt{\lambda_{\max}(G^*(j\omega)G(j\omega))}\notag \\
	=& \sup_\omega \max_\ell \frac{1}{\sqrt{(-\omega^2+\lambda_\ell k_0)^2+b_0^2\omega^2\lambda_\ell^2 }}\notag\\
	=&  \max_\ell \max_\omega |G_\ell(j\omega)|= \max_\ell A_\ell =A_1.
\end{align*}
where $A_1$ is given in~\eqref{eq:amplitude2}. 

To prove the second part of the theorem, we  rewrite the coupled-ODE model~\eqref{eq:coupled_ODE} as
\begin{align}\label{eq:expansion}
	\ddot{\tilde{p}}+b_0 L\dot{\tilde{p}}+k_0 L \tilde{p}=v_1\sin(\omega_1 t). 
\end{align}
By the method of eigenfuction expansion~\cite{haberman}, we can write $\tilde{p}(t)=\sum_{\ell=1}^N v_\ell h_\ell(t)$, where $v_\ell$'s are the eigenvectors of $L_g^{(u)}$. Substituting it into Eq.~\eqref{eq:expansion}, we obtain 
\begin{align*}
 \sum_{\ell=1}^N (v_\ell \ddot{h}_\ell(t)+b_0 Lv_\ell \dot{h}_\ell(t)+k_0 L v_\ell h_\ell(t))=v_1\sin(\omega_1 t). 
\end{align*}
Due to superposition property of linear system, the above equation can be split into $N$ ordinary differential equations by using $L v_\ell = \lambda_\ell v_\ell$,
\begin{align*}
\ddot{h}_1(t)+b_0\lambda_\ell \dot{h}_1(t)+k_0\lambda_\ell h_1(t)&=\sin(\omega_1 t), \\
\ddot{h}_\ell(t)+b_0 \lambda_\ell \dot{h}_\ell(t)+k_0 \lambda_\ell h_\ell(t)&=0, \quad \ell \in \{2,\cdots, N\}.
\end{align*}
Following straightforward algebra, the steady-state response of each $h_\ell(t)$ is given by
\begin{align*}
h_1(t)&=A_1 \sin(\omega_1 t+\psi_1),\\
h_\ell(t)&=0, \quad \ell \in \{2,\cdots, N\},
\end{align*}
where $A_1$ is given in~\eqref{eq:amplitude2}. Thus the steady state response of $\tilde{p}$ is given by $\tilde{p}=v_1 A_1 \sin(\omega_1 t+\psi_1)$, which yields 
\begin{align*}
	\frac{\|\tilde{p}\|_{\L_2}}{\|w\|_{\L_2}}=A_1 =	AF^{linear}. \tag*{\frQED}
\end{align*}
\end{proof-theorem}

\begin{corollary}\label{cor:bidi}
Consider an $N$-vehicle platoon with symmetric bidirectional architecture. With linear controller $f(x)=k_0x$ and $g(x)=b_0x$ in~\eqref{eq:u-bidi}, the  amplification factor  satisfies 
\begin{align*}
\Big(	\frac{1}{b_0\sqrt{k_0}\pi^3}  \Big) (2N+1)^3 &\leq 	AF^{linear} \leq \Big(\frac{1}{4b_0\sqrt{2k_0}}\Big) (2N+1)^3.
 \end{align*}
Furthermore, when $N \gg 1$, the amplification factor and its corresponding peak frequency are asymptotically
\begin{align*}
	AF^{linear} \approx \frac{8N^3}{\sqrt{k_0}b_0\pi^3},\quad \omega_p=\omega_1 \approx \frac{\sqrt{k_0}\pi}{2N}.\tag*{\frqed}
\end{align*}
\end{corollary}

\begin{proof-corollary}{~\ref{cor:bidi}}
	Following Theorem 2 of~\cite{yueh_tridiag}, the eigenvalues of the grounded graph Laplacian $L_g^{(u)}$ (given in~\eqref{eq:redu_laplacian}) of a 1-D vehiculear platoon  are given by
\begin{align*}
	\lambda_1&=2-2\cos(\frac{\pi}{2N+1})=4\sin^2(\frac{\pi}{2(2N+1)}).
\end{align*}
It follows from the fact $\frac{2}{\pi}x \leq \sin x \leq x, \forall x \in [0,\frac{\pi}{2}]$ that  
\begin{align}\label{eq:lambda_inequality}
 \frac{4}{(2N+1)^2} \leq \lambda_\ell \leq \frac{\pi^2}{(2N+1)^2}.
\end{align}
For any fixed $k_0, b_0$, when $N$ is large, we have $\lambda_1 <2k_0/b_0^2$, which implies 
\begin{align*}
	A_1=\frac{2}{\lambda_1^{3/2}b_0\sqrt{4k_0-\lambda_1b_0^2}}\geq  \frac{1}{\lambda_1^{3/2}b_0\sqrt{k_0}} \geq  \frac{1}{b_0\sqrt{k_0}} \frac{(2N+1)^3}{\pi^3},
\end{align*}
where the last inequality is obtained from~\eqref{eq:lambda_inequality}.  In addition, using $\lambda_1 <2k_0/b_0^2$ and  $\frac{1}{\lambda_\ell^{3/2}} \leq \frac{(2N+1)^3}{8}$, we obtain an upper bound  
\begin{align*}
A_1=\frac{2}{\lambda_1^{3/2}b_0\sqrt{4k_0-\lambda_1b_0^2}} \leq \frac{2}{\lambda_1^{3/2}b_0\sqrt{2k_0}} \leq \frac{(2N+1)^3}{4b_0\sqrt{2k_0}}.
\end{align*}

To get the asynptotic formula, when $N$ is large, we use the approximation $\lambda_1 \approx \frac{\pi^2}{4N^2}$. Therefore, $\lambda_1 < 2k_0/b_0^2$ is true for large enough $N$ irrespective of the values of $k_0$ and $b_0$. Substituting $\lambda_1\approx \frac{\pi^2}{4N^2}$ into~\eqref{eq:amplitude2} and~\eqref{eq:frequency2}, we obtain that
\begin{align*}
	AF^{linear}=A_1 \approx  \frac{8N^3}{\sqrt{k_0}b_0\pi^3}, \quad \omega_p=\omega_1 \approx \frac{\sqrt{k_0}\pi}{2N}, 
\end{align*}
which completes the proof. \frQED
\end{proof-corollary}

\subsection{Numerical comparison of sensitivity to disturbance between linear and non-linear controllers}
In this section, we present results of  numerical computations in order to obtain the sensitivity to external disturbances of the 1-D vehicular platoon with nonlinear controllers. These are then compared against the predictions on similar performance measures for the linear controllers that have been analytically obtained in the previous section. The performance measures are also numerically computed for the linear controllers in order to verify the analytical predictions.  The controllers used are the ones given by~\eqref{eq:control_function} in Section~\ref{sec:conv_rate}.  

\begin{figure}
	  \psfrag{N}{\scriptsize $N$}
	  \psfrag{amplificationfactor}{\scriptsize $\quad AF,\quad  \hat{AF}$}
	  \psfrag{linear}{\scriptsize Linear (simulation)}
	  \psfrag{analytical1}{\scriptsize Asymptotic formula of Cor.~\ref{cor:pre}}
	  \psfrag{analytical}{\scriptsize Asymptotic formula of Cor.~\ref{cor:bidi}}
	  \psfrag{nonlinear}{\scriptsize Nonlinear (simulation)}
	  \psfrag{LB}{\scriptsize Lower bound (Corollary~\ref{cor:pre})}
	  \psfrag{UB}{\scriptsize Upper bound (Corollary~\ref{cor:pre})}
          \psfrag{LB1}{\scriptsize Lower bound (Corollary~\ref{cor:bidi})}
	  \psfrag{UB1}{\scriptsize Upper bound (Corollary~\ref{cor:bidi})}
\centering
\subfigure[Predecessor-following]{\includegraphics[scale = 0.34]{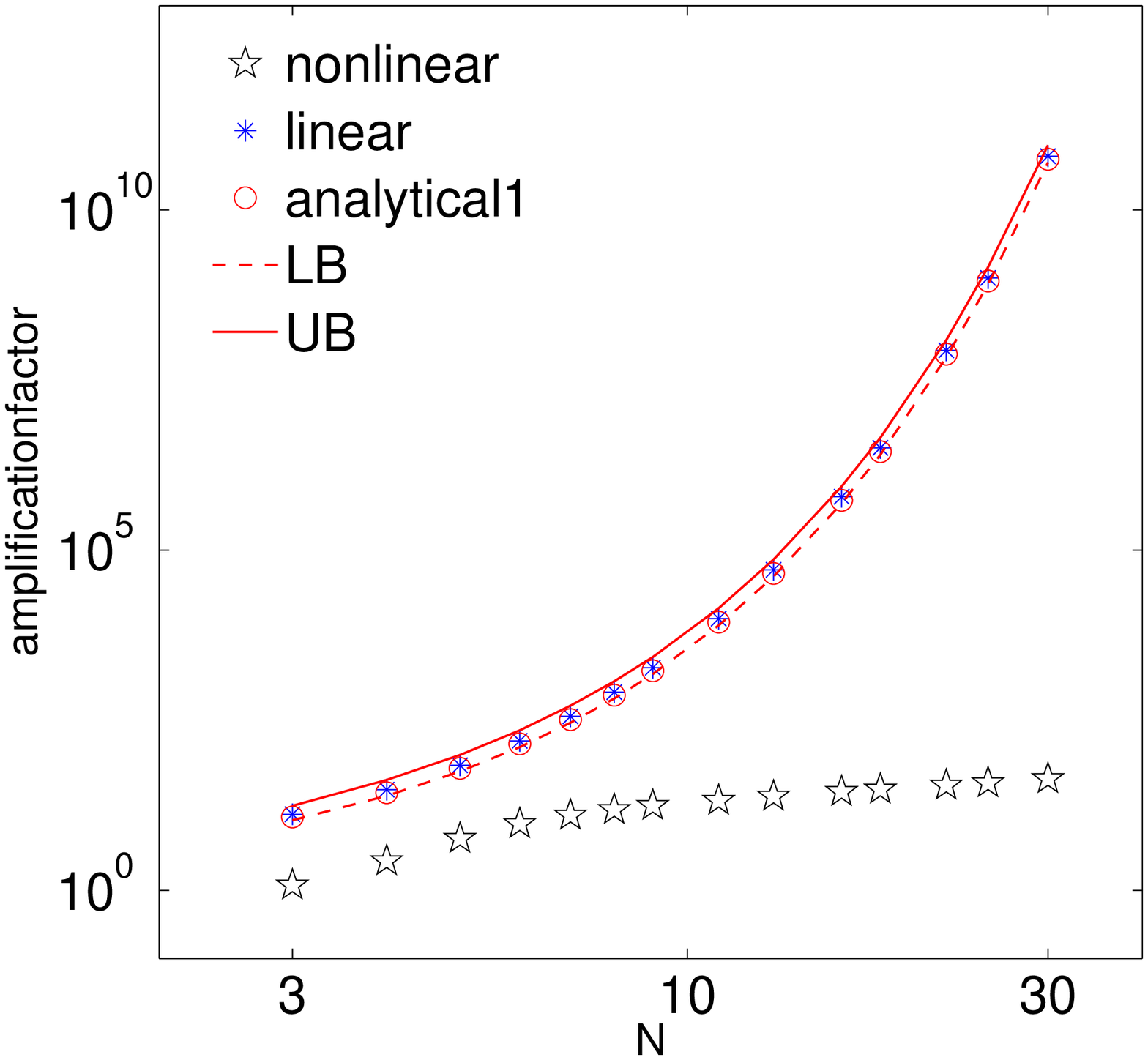} } 
\subfigure[Symmetric bidirectional]{\includegraphics[scale = 0.34]{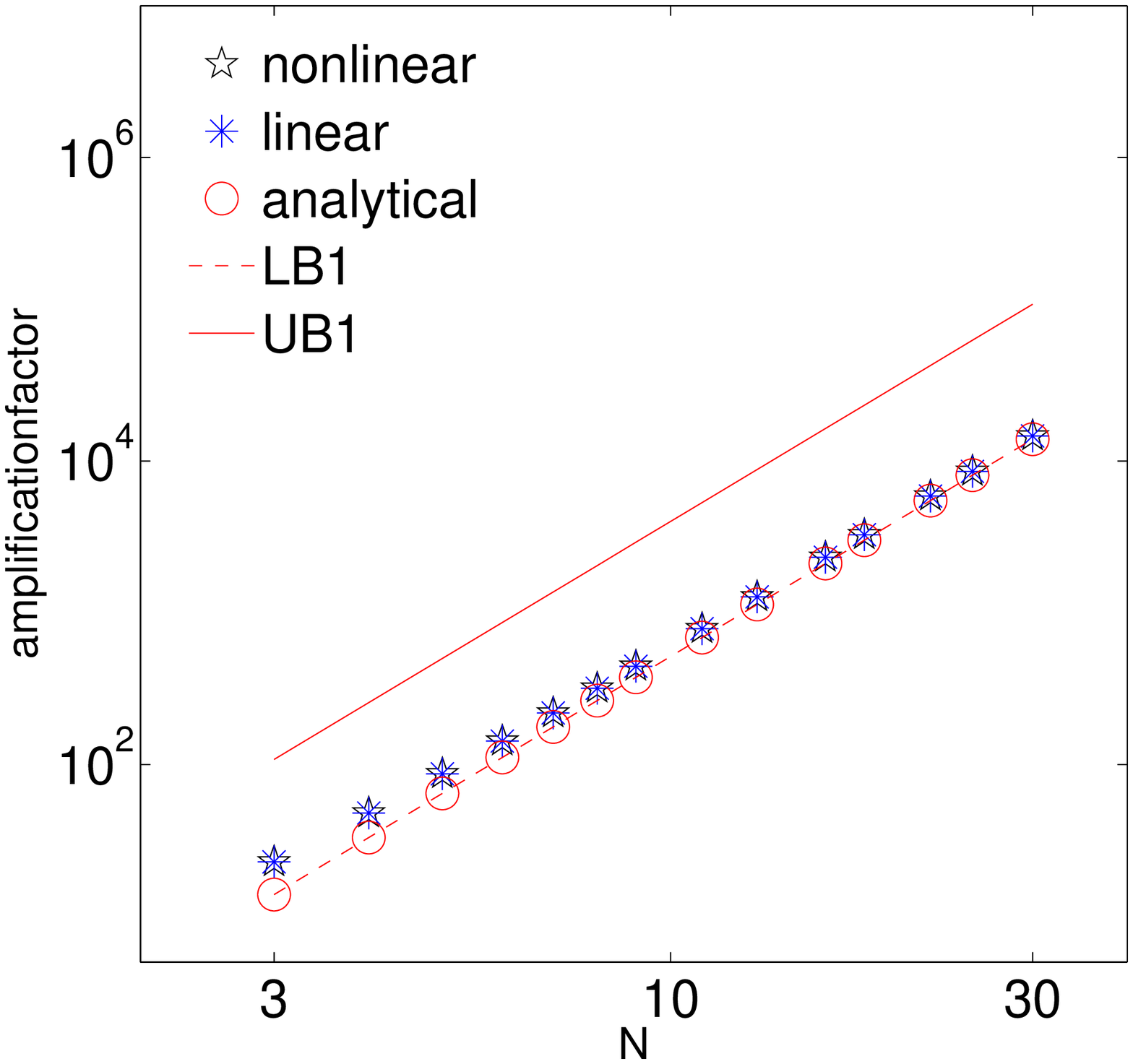}}
\caption{Comparison of  amplification factor of the platoon with linear and nonlinear controller for both the predecessor-following and symmetric bidirectional architectures.
}
\label{fig:l2_all}
\end{figure}

Figure~\ref{fig:l2_all} shows the amplification factor as a function of $N$. The following observations are obtained: 1) The lower and upper bounds and asymptotically formulae derived in Corollary~\ref{cor:pre} and Corollary~\ref{cor:bidi} are quite accurate; 2) Comparing Figure~\ref{fig:l2_all} (a) and (b) we see that the symmetric bidirectional architecture has a smaller amplification factor than the predecessor-following case when the controller is linear.  However, when nonlinear controller is applied, the symmetric bidirectional architecture has a worse scaling trend than that of the predecessor-following case;  3) In the predecessor-following architecture, the growth of the amplification factor with respect to $N$ is much slower with the nonlinear controller than with the linear controller.  In the symmetric bidirectional architecture, there is little difference between the two controllers for this sinusoidal disturbance.

\begin{figure}
	  \psfrag{N}{\scriptsize $N$}
	  \psfrag{ATA Ratio}{\scriptsize $\quad R,\ \hat{R}$}
	  \psfrag{linear}{\scriptsize Linear (simulation)}
	  \psfrag{nonlinear}{\scriptsize Nonlinear (simulation)}
  	  \psfrag{analytical}{\scriptsize Linear (Eq.~\eqref{eq:lya_equation})}
\centering
\subfigure[Predecessor-following]{\includegraphics[scale = 0.34]{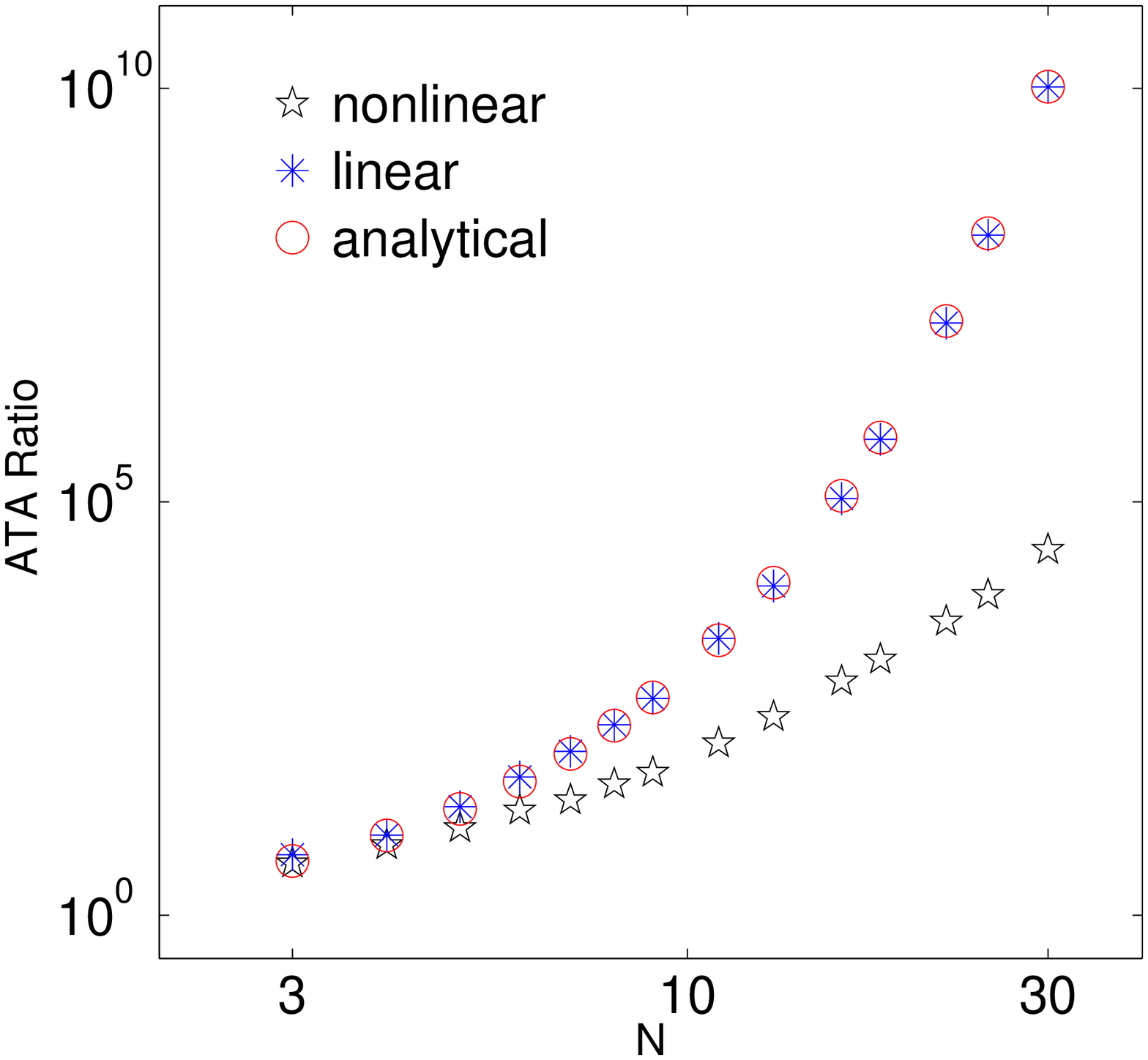} }
\subfigure[Symmetric bidirectional]{\includegraphics[scale = 0.34]{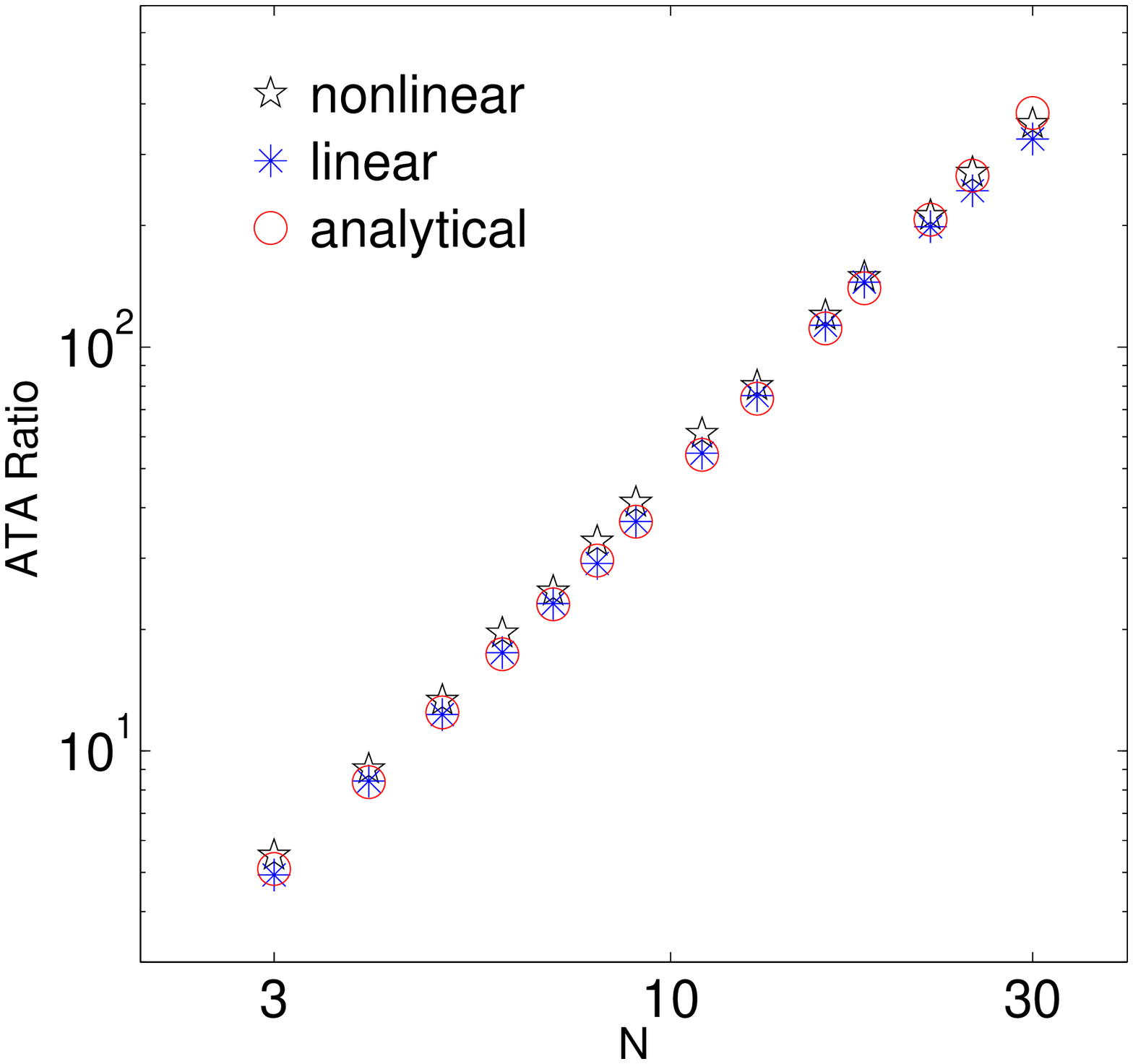}}

\caption{Comparison of the ratios $R,\ \hat{R}$ as a function of the number of vehicles $N$ with white noise disturbances. The value of $\sigma_0$ used is $1$.}
\label{fig:noise_N}
\end{figure}

\begin{figure}
	  \psfrag{A}{\scriptsize $\sigma_0$}
	  \psfrag{ATA Ratio}{\scriptsize $\quad  R,\ \hat{R}$}
	  \psfrag{linear}{\scriptsize Linear (simulation)}
	  \psfrag{nonlinear}{\scriptsize Nonlinear (simulation)}
  	  \psfrag{analytical}{\scriptsize Linear (Eq.~\eqref{eq:lya_equation})}
\centering
\subfigure[Predecessor-following]{\includegraphics[scale = 0.34]{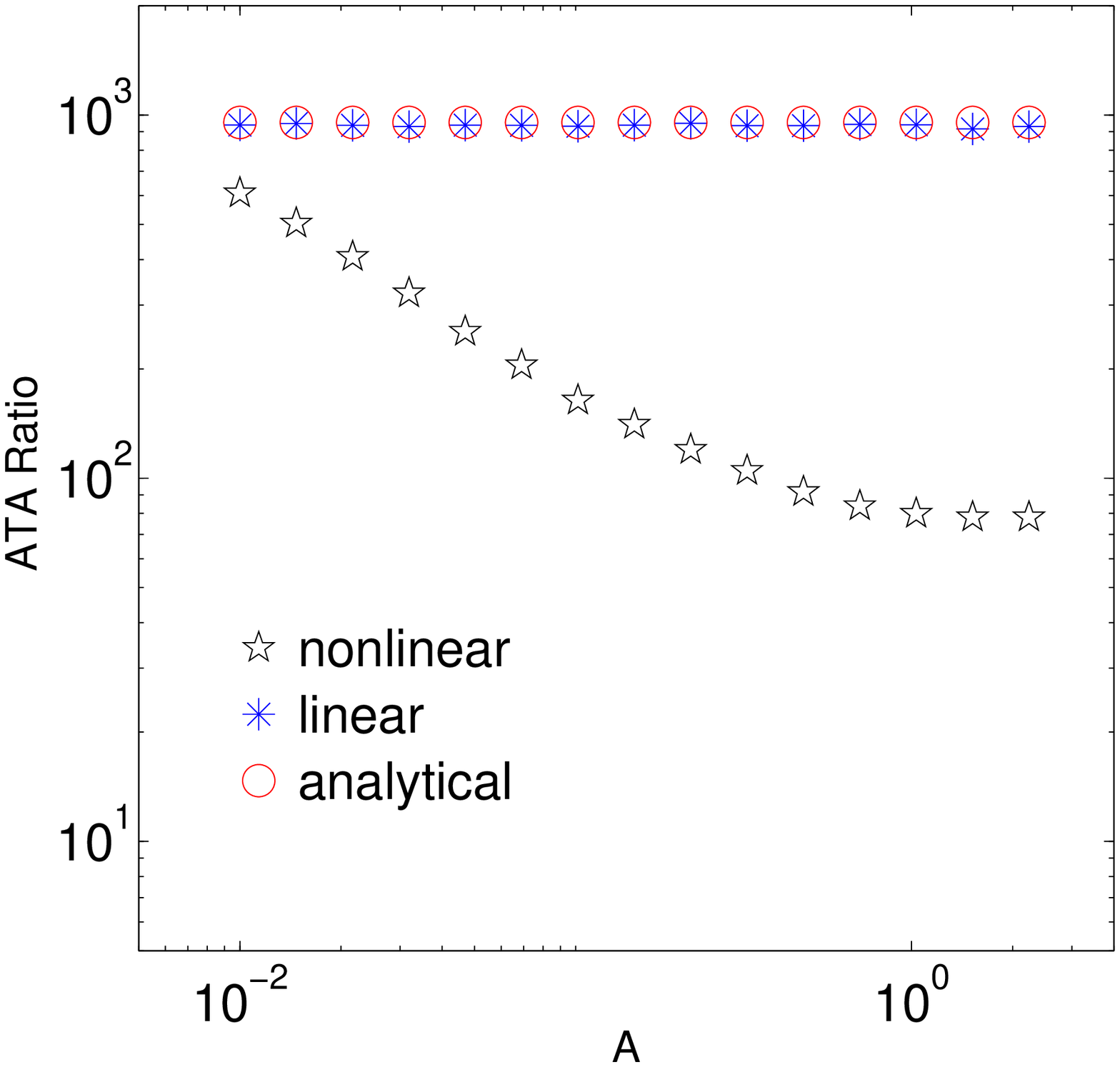} }
\subfigure[Symmetric bidirectional]{\includegraphics[scale = 0.34]{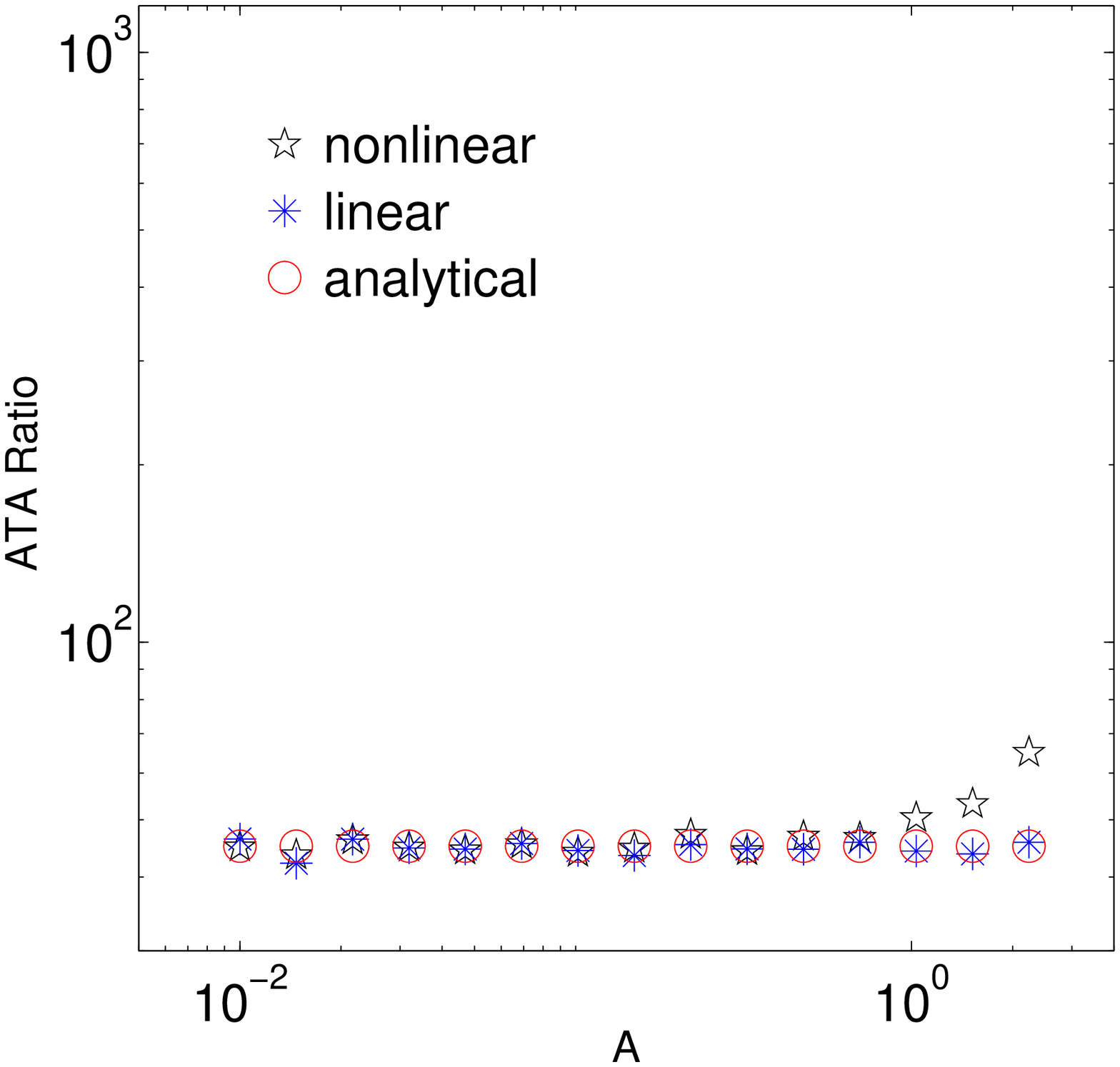}}

\caption{Comparison of the ratios $R,\ \hat{R}$ of a $10$-vehicle platoon as a function of the standard deviation $\sigma_0$ of the white noises.}
\label{fig:noise_amplitude}
\end{figure}

To examine the effect of random disturbances, we compute the estimate of $R$ that is defined in~\eqref{eq:R-def} for $T=3000$ seconds, through Monte-Carlo simulations for both linear and non-linear cases. Figure~\ref{fig:noise_N} shows $\hat{R}$ vs. $N$ for a fixed $\sigma_0$, the strength of the white noise,  while Figure~\ref{fig:noise_amplitude} shows $\hat{R}$ vs. $\sigma_0$ for a fixed $N$. The following broad conclusions can be made from these simulations: 1) disturbance amplification is far worse in the predecessor following architecture than in the symmetric bidirectional one, and this conclusion holds whether one uses a linear or non-linear control; compare Figure~\ref{fig:noise_N} (b) vs. Figure~\ref{fig:noise_N} (a) and Figure~\ref{fig:noise_amplitude} (b) vs. Figure~\ref{fig:noise_amplitude} (a). This shows that architecture has a more profound impact on performance than linearity or non-linearity, and given a choice, symmetric bidirectional is preferable. Though these conclusions are for a specific non-linear control design, simulations with other non-linear control laws that are not reported here due to lack of space shows the same trend. These laws include (a) $f(z) = z^3$, $g(z) = bz$, and (b) $f(z) = \sat(z)$, $g(z) = bz$; 2) If symmetric bidirectional architecture is indeed used, both the linear and non-linear control laws have almost identical robustness  to disturbances. The only exception is when the strength of the white noise $\sigma_0$ is large, in which case the non-linear control law performs poorly compared to the linear one; 3) If the predecessor architecture is to be used due to other constraints, the non-linear control law is preferable since it has better robustness to disturbance than its linear counterpart; see Figure~\ref{fig:noise_N} (a)  and Figure~\ref{fig:noise_amplitude} (a).

\section{Conclusion}\label{sec:conc}
 We studied the stability and robustness of large vehicular formations with both linear and a class of nonlinear decentralized controllers. Under certain sector assumptions on the non-nonlinear control gain functions, we proved the stability of the formations with both the directed tree and undirected graphs with spanning trees. In the case of linear control, we analyzed the transient response due to initial conditions in terms of stability margin and the multiplicity of the least stable eigenvalues. We showed that, on one hand, although the formation with directed tree graph achieves a size-independent stability margin, it suffers from high algebraic growth of initial conditions because of the high multiplicity of its least stable eigenvalue. On the other hand, the stability margin of the formation with undirected information graph decays to $0$ at least as $O(1/N)$ if it has bounded degree. Moreover, we derived the scaling laws of the amplification factor  with respect to the size of the formation. In particular, for 1-D vehicular platoons, we show that the amplification factor scales as $O(\alpha^N)$ ($\alpha>1$) for the predecessor-following architecture but only as $O(N^3)$ for the symmetric bidirectional architecture. We examined the transient performance and sensitivity to external (sinusoidal or random) disturbances with nonlinear controllers by extensive numerical simulations. These simulations show that in case of the directed tree graphs, a class of nonlinear controllers perform better compared to the linear control, both in terms of transient performance and sensitivity to external disturbances. In the case of undirected information graphs, numerical simulations indicate that the nonlinear controller yields equal or worse performance compared to the linear one.

 We have assumed that each vehicle has the same open-loop dynamics and uses the same control law. In the linear symmetric bidirectional case, it was shown in~\cite{HH_PB_CDC:10} that heterogeneity in vehicle masses and control gains has little effect on the stability margins. The non-linear stability analysis in this paper can be extended to the case when vehicles have different masses and employs a different controller in a straightforward manner. However, lack of symmetry has a large effect, as the difference between directed tree and undirected graphs established here shows. In the case of 1-D vehicular platoon with linear control, a general asymmetric architecture that lies in between symmetric bidirectional and predecessor-following architectures has been examined in~\cite{HH_PB_CDC:10} and~\cite{veerman_stability}. It was shown in~\cite{HH_PB_CDC:10} that such asymmetry can lead to improvement in the stability margin. Numerical simulations in~\cite{HH_PB_relative_vel_arxiv} indicate that if the asymmetry is only in the velocity feedback terms, sensitivity to external disturbances is improved as well. In contrast, it was shown in~\cite{veerman_stability} that if equal asymmetry is applied to both position and velocity feedback terms, sensitivity to external disturbances worsens. Analysis of  stability with non-linear asymmetric control and sensitivity to disturbance with asymmetric control (linear or non-linear) are open problems.

\section*{Appendix}
\begin{proof-proposition}{\ref{prop:2nd-order-stab}}
First, we consider the unforced system with state $y=[y_1,y_2]^T$, whose dynamics are
\begin{equation}\label{eq:ss1}
  \begin{split}
	\dot{y}_1&=y_2, \\
	\dot{y}_2&=-f(y_1)- g(y_2).    
  \end{split}
\end{equation}
Consider the following Lyapunov function candidate:
\begin{align}\label{eq:lyap_f}
	V(y)=\frac{1}{2} y^{T} P y + \gamma \int_0^{y_1} f(z)  dz,
\end{align}
where $\displaystyle{P=\begin{bmatrix}
	1 & 1 \\ 1 & \gamma 
\end{bmatrix}}$ and $\gamma\geq \max{\{ 1, \frac{1}{\varepsilon_2}+\frac{(1+K_2)^2}{\varepsilon_1 \varepsilon_2}}\}$, which ensures that $P$ is positive definite. From the Rayleigh Ritz Theorem~\cite{khalil}, we have the following inequality $\lambda_{\min}(P) \|y\|^2 \leq y^T P y \leq \lambda_{\max}(P) \|y\|^2$, where $\lambda_{\min}(P)>0, \lambda_{\max}(P)>0$ are the minimum and maximum eigenvalues of $P$ respectively. This shows that $V(y)$ is radially unbounded, and in addition satisfies the following inequality
\begin{align}\label{eq:V_bound}
	V(y) \leq \frac{\lambda_{\max}(P)}{2} \|y\|^2 + \gamma \int_0^{y_1} f(z)  dz \leq \frac{\lambda_{\max}(P)}{2} \|y\|^2 + \frac{\gamma K_1}{2} y_1^2 \leq \frac{\lambda_{\max}(P)+\gamma K_1}{2} \|y\|^2, 
\end{align}
where the second inequality follows from the fact that the function $f(z)$ belongs to the sector $[\varepsilon_1, K_1]$. The derivative of $V$ along the trajectory of~\eqref{eq:ss1} is given by
\begin{align}\label{eq:lyap_der}
	\dot{V}&=y^T P \dot{y} +\gamma f(y_1) y_2 \notag\\
	      &= -y_1 f(y_1) -\gamma y_2g(y_2) +y_2^2 +y_1y_2-y_1g(y_2) \notag \\
	      &\leq -\varepsilon_1 y_1^2 -(\gamma \varepsilon_2-1) y_2^2+ (1+K_2) |y_1||y_2|,\notag \\
	      & \leq  -\frac{1}{2} (\varepsilon_1 y_1^2 +(\gamma \varepsilon_2-1) y_2^2) -\frac{1}{2} [\varepsilon_1 y_1^2 -2(1+K_2) |y_1||y_2|+(\gamma \varepsilon_2-1) y_2^2)] \notag \\
	      &\leq  -\frac{1}{2} (\varepsilon_1 y_1^2 +(\gamma \varepsilon_2-1) y_2^2)\leq -\frac{1}{2} \min\{{\varepsilon_1, (\gamma\varepsilon_2-1)\}}\|y\|^2,
\end{align}
where the second last inequality following from $\gamma\geq \max{\{ 1, \frac{1}{\varepsilon_2}+\frac{(1+K_2)^2}{\varepsilon_1 \varepsilon_2}}\}$, since that leads to the conclusion, upon a completion of squares, that $\varepsilon_1y_1^2 - 2(1+K_2)|y_1||y_2| + (\gamma\varepsilon_2-1)y_2^2 \geq 0$. Since $V$ is radially unbounded and satisfies~\eqref{eq:V_bound}, it follows from~\eqref{eq:lyap_der} that the origin $y=0$ of~\eqref{eq:ss1} is globally exponentially stable. Since the functions $f, g$ are assumed to be smooth enough, the ISS property follows from the fact that a globally exponentially stable system with input $u$ is ISS~\cite[Lemma 4.6]{khalil}. \frQED
\end{proof-proposition}

\bibliographystyle{IEEEtran}
\bibliography{../../PBbib/HH}

\begin{thebibliography}{10}
\providecommand{\url}[1]{#1}
\csname url@rmstyle\endcsname
\providecommand{\newblock}{\relax}
\providecommand{\bibinfo}[2]{#2}
\providecommand\BIBentrySTDinterwordspacing{\spaceskip=0pt\relax}
\providecommand\BIBentryALTinterwordstretchfactor{4}
\providecommand\BIBentryALTinterwordspacing{\spaceskip=\fontdimen2\font plus
\BIBentryALTinterwordstretchfactor\fontdimen3\font minus
  \fontdimen4\font\relax}
\providecommand\BIBforeignlanguage[2]{{%
\expandafter\ifx\csname l@#1\endcsname\relax
\typeout{** WARNING: IEEEtran.bst: No hyphenation pattern has been}%
\typeout{** loaded for the language `#1'. Using the pattern for}%
\typeout{** the default language instead.}%
\else
\language=\csname l@#1\endcsname
\fi
#2}}

\bibitem{JH_MT_PV_CSM:94}
J.~K. Hedrick, M.~Tomizuka, and P.~Varaiya, ``Control issues in automated
  highway systems,'' \emph{{IEEE} Control Systems Magazine}, vol.~14, pp. 21 --
  32, December 1994.

\bibitem{okubo_flock}
A.~Okubo, ``{Dynamical aspects of animal grouping: swarms, schools, flocks, and
  herds},'' \emph{Advances in Biophysics}, vol.~22, pp. 1--94, 1986.

\bibitem{tanner2007decentralized}
H.~Tanner and D.~Christodoulakis, ``{Decentralized cooperative control of
  heterogeneous vehicle groups},'' \emph{Robotics and autonomous systems},
  vol.~55, no.~11, pp. 811--823, 2007.

\bibitem{DJ_WB_MP_EW_AIAA:02}
E.~Wagner, D.~Jacques, W.~Blake, and M.~Pachter, ``Flight test results of close
  formation flight for fuel savings,'' in \emph{{AIAA} Atmospheric Flight
  Mechanics Conference and Exhibit}, 2002, {AIAA}-2002-4490.

\bibitem{minesweeping}
P.~M. Ludwig, ``Formation control for multi-vehicle robotic minesweeping,''
  Master's thesis, Naval postgraduate school, 2000.

\bibitem{stringstability:96}
S.~Darbha and J.~K. Hedrick, ``String stability of interconnected systems,''
  \emph{IEEE Transactions on Automatic Control}, vol.~41, no.~3, pp. 349--356,
  March 1996.

\bibitem{Seiler_disturb_TAC:04}
P.~Seiler, A.~Pant, and J.~K. Hedrick, ``Disturbance propagation in vehicle
  strings,'' \emph{IEEE Transactions on Automatic Control}, vol.~49, pp.
  1835--1841, October 2004.

\bibitem{zhang1999using}
Y.~Zhang, B.~Kosmatopoulos, P.~Ioannou, and C.~Chien, ``{Using front and back
  information for tight vehicle following maneuvers},'' \emph{Vehicular
  Technology, IEEE Transactions on}, vol.~48, no.~1, pp. 319--328, 1999.

\bibitem{darbha1994comparison}
S.~Darbha, J.~Hedrick, C.~Chien, and P.~Ioannou, ``{A comparison of spacing and
  headway control laws for automatically controlled vehicles},'' \emph{Vehicle
  System Dynamics}, vol.~23, no.~8, pp. 597--625, 1994.

\bibitem{peppard_string}
L.~Peppard, ``{String stability of relative-motion PID vehicle control
  systems},'' \emph{Automatic Control, IEEE Transactions on}, vol.~19, no.~5,
  pp. 579--581, 1974.

\bibitem{chu_platoon}
K.~Chu, ``{Decentralized control of high-speed vehicular strings},''
  \emph{Transportation Science}, vol.~8, no.~4, p. 361, 1974.

\bibitem{klinge_string:09}
S.~Klinge and R.~H. Middleton, ``{String stability analysis of homogeneous
  linear unidirectionally connected systems with nonzero initial conditions},''
  in \emph{Signals and Systems Conference ({ISSC} 2009)}.\hskip 1em plus 0.5em
  minus 0.4em\relax IET, 2009, pp. 1--6.

\bibitem{Khatir}
M.~E. Khatir and E.~J. Davison, ``Decentralized control of a large platoon of
  vehicles using non-identical controllers,'' in \emph{Proceedings of the 2004
  American Control Conference}, 2004, pp. 2769--2776.

\bibitem{PB_JH_CDC:05}
P.~Barooah and J.~Hespanha, ``{Error amplification and disturbance propagation
  in vehicle strings with decentralized linear control},'' in \emph{44th {IEEE}
  Conference on Decision and Control}.\hskip 1em plus 0.5em minus 0.4em\relax
  IEEE, 2005, pp. 4964 -- 4969.

\bibitem{lestas_scalability}
I.~Lestas and G.~Vinnicombe, ``{Scalability in heterogeneous vehicle
  platoons},'' in \emph{American Control Conference, 2007. ACC'07}.\hskip 1em
  plus 0.5em minus 0.4em\relax IEEE, 2007, pp. 4678--4683.

\bibitem{AP_PS_KH_TAC:02}
A.~Pant, P.~Seiler, , and K.~Hedrick, ``Mesh stability of look-ahead
  interconnected systems,'' \emph{{IEEE} Transactions on Automatic Control},
  vol.~47, pp. 403--407, February 2002.

\bibitem{SKY_SD_KRR_TAC:06}
S.~K. Yadlapalli, S.~Darbha, and K.~R. Rajagopal, ``Information flow and its
  relation to stability of the motion of vehicles in a rigid formation,''
  \emph{{IEEE} Transactions on Automatic Control}, vol.~51, no.~8, August 2006.

\bibitem{SD_PP_IJES:10}
S.~Darbha and P.~R. Pagilla, ``Limitations of employing undirected information
  flow graphs for the maintenance of rigid formations for heterogeneous
  vehicles,'' \emph{International journal of engineering science}, vol.~48,
  no.~11, pp. 1164--1178, 2010.

\bibitem{feedback_linearization}
S.~Stankovic, M.~Stanojevic, and D.~Siljak, ``{Decentralized overlapping
  control of a platoon of vehicles},'' \emph{Control Systems Technology, IEEE
  Transactions on}, vol.~8, no.~5, pp. 816--832, 2000.

\bibitem{munz2011robust}
U.~Munz, A.~Papachristodoulou, and F.~Allgower, ``{Robust Consensus Controller
  Design for Nonlinear Relative Degree Two Multi-Agent Systems With
  Communication Constraints},'' \emph{Automatic Control, IEEE Transactions on},
  vol.~56, no.~1, pp. 145--151, 2011.

\bibitem{veerman_stability}
J.~Veerman, ``Stability of large flocks: an example,'' July 2009,
  {arXiv}:1002.0768.

\bibitem{bamjovmitCDC08}
B.~Bamieh, M.~R. Jovanovi\'c, P.~Mitra, and S.~Patterson, ``Effect of
  topological dimension on rigidity of vehicle formations: fundamental
  limitations of local feedback,'' in \emph{Proceedings of the 47th IEEE
  Conference on Decision and Control}, Cancun, Mexico, 2008, pp. 369--374.

\bibitem{HH_PB_PM_TAC:11}
H.~Hao, P.~Barooah, and P.~G. Mehta, ``Stability margin scaling of distributed
  formation control as a function of network structure,'' \emph{IEEE
  Transactions on Automatic Control}, vol.~56, no.~4, pp. 923--929, April 2011.

\bibitem{barooah2006graph}
P.~Barooah and J.~Hespanha, ``Graph effective resistance and distributed
  control: Spectral properties and applications,'' in \emph{Decision and
  Control, 2006 45th IEEE Conference on}.\hskip 1em plus 0.5em minus
  0.4em\relax IEEE, 2006, pp. 3479--3485.

\bibitem{SKY_MSThesis}
S.~K. Yadlapalli, ``Information flow and its relation to the stability of
  controlled motion of vehicles in a rigid formation,'' Master's thesis, Texas
  A\&M University, 2005.

\bibitem{yueh_tridiag}
W.~Yueh, ``{Eigenvalues of several tridiagonal matrices},'' \emph{Applied
  Mathematics E-Notes}, vol.~5, pp. 66--74, 2005.

\bibitem{khalil}
H.~Khalil, \emph{{Nonlinear Systems 3rd}}.\hskip 1em plus 0.5em minus
  0.4em\relax Prentice hall Englewood Cliffs, NJ, 2002.

\bibitem{simon_estimation}
D.~Simon, \emph{{Optimal state estimation: Kalman, H [infinity] and nonlinear
  approaches}}.\hskip 1em plus 0.5em minus 0.4em\relax John Wiley and Sons,
  2006.

\bibitem{SDE}
D.~Higham, ``{An algorithmic introduction to numerical simulation of stochastic
  differential equations},'' \emph{SIAM review}, vol.~43, no.~3, pp. 525--546,
  2001.

\bibitem{matrix_cookbook}
K.~Petersen and M.~Pedersen, ``{The matrix cookbook},'' \emph{Technical
  University of Denmark}, 2006.

\bibitem{h8norm}
G.~Balas and A.~Packard, ``{The structured singular value ($\mu$) framework},''
  \emph{The Control Handbook}, pp. 671--687.

\bibitem{haberman}
R.~Haberman, \emph{{Elementary applied partial differential equations: with
  Fourier series and boundary value problems}}.\hskip 1em plus 0.5em minus
  0.4em\relax Prentice-Hall, 2003.

\bibitem{HH_PB_CDC:10}
H.~Hao and P.~Barooah, ``Control of large 1d networks of double integrator
  agents: role of heterogeneity and asymmetry on stability margin,'' in
  \emph{{IEEE} Conference on Decision and Control}, December 2010, pp. 7395 --
  7400.

\bibitem{HH_PB_relative_vel_arxiv}
------, ``{Control of large 1D networks of double integrator agents: role of
  heterogeneity and asymmetry on stability margin},'' \emph{Arxiv preprint
  arXiv:1011.0791}, 2010.

\end{thebibliography}

\end{document}